# Pay and Beyond: The Evolving Rewards for AI Skills

Alejandra Castaneda[a], Matthew Bone[a,b], and Fabian Stephany📧[a,c,d]

a) Oxford Internet Institute, University of Oxford, UK, b) Burning Glass Institute, USA, c) Institute for New Economic Thinking, Oxford Martin School, d) Bruegel, Brussels, Belgium.

📧 fabian.stephany@oii.ox.ac.uk

October 2026

**Abstract**

This study examines the monetary and non-monetary rewards advertised in job vacancies requiring artificial intelligence skills in the U.S. labor market. Using approximately ten million online job vacancies posted between 2018 and 2025, we examine the development of AI demand, wage premia, and non-monetary benefits — including tuition assistance, paid leave, health and well-being programs, parental leave, inclusive workplace language, and remote work — across firms of different sizes. We find that demand for AI skills increased sharply following the introduction of generative AI in late 2022 across small and medium-sized enterprises (SMEs), large firms, and S&P 500 firms, while AI wage premia rose most strongly among SMEs. AI vacancies are significantly more likely than comparable non-AI roles to advertise non-monetary benefits, particularly remote work, inclusive culture, and parental leave. This premium is strongest among SMEs and large firms, but weaker among S&P 500 firms. Finally, we show that the relationship between wages and benefits shifted after the spread of generative AI: whereas complementarity between high wages and benefits was previously concentrated among S&P 500 firms, and substitution among SMEs, the difference across firm sizes has attenuated. Our findings indicate that the increased demand for AI skills has been accompanied by changes in compensation patterns beyond pay, as employers across the size distribution combine monetary and non-monetary rewards to recruit AI-skilled workers.

*Keywords*: Future of Work, Labor Markets, Skills, AI, Automation

*JEL Class*: C55, I23, J23, J24, J31

## 1. Introduction

With rapidly advancing technology, and particularly the diffusion of artificial intelligence (AI), the nature of jobs and the demand for skills are changing rapidly and becoming increasingly difficult to meet (Acemoglu et al., 2022; Alekseeva et al., 2021; Devigili et al., 2026; OECD, 2026; Tamayo et al., 2023). At the same time, new generations of workers are entering the labor market with different expectations about what employers should offer, especially in terms of flexibility, inclusive culture, and well-being (Rodríguez-Sánchez et al., 2020). One of the clearest signs of the current shift in labor market dynamics is the growing demand for AI talent. With the rise of generative AI and machine learning, there has been exponential growth in AI-related roles across industries. However, the supply of qualified professionals has not kept pace, creating a bottleneck in workforce development. Prior research has shown that AI roles command substantial wage premia (Alekseeva et al., 2021; Bone et al., 2025; Green & Lamby, 2023).

Yet salary is only one dimension of how employers compete for talent. This paper investigates whether firms also use non-monetary benefits — such as paid leave, tuition assistance, workplace wellness programs, parental leave, remote work, and inclusive work cultures — in recruiting AI talent. Specifically, we ask: ***Do firms offer non-monetary benefits to attract AI talent***? And if so, how do firms combine wages and non-monetary benefits when recruiting for scarce AI skills? To explore these questions, we analyze a comprehensive dataset of nearly ten million online job vacancies in the U.S. between January 2018 and December 2025. These job postings include detailed information about required skills, job characteristics, and — critically — whether specific non-monetary benefits are explicitly mentioned in the advertisement.

Our findings reveal several key insights. First, AI-related vacancies are significantly more likely to advertise non-monetary benefits than comparable non-AI roles, particularly parental leave, inclusive culture, and remote work. These relationships remain robust after controlling for industry, occupation, education requirements, experience, geography, and time, although we find little evidence for more standardized or costly benefits such as paid leave and tuition assistance. Second, we show that the rapid increase in demand for AI skills following the introduction of generative AI has occurred across firms of all sizes, while the strongest growth in AI wage premia is observed among small and medium-sized enterprises (SMEs). The non-monetary premium for AI roles also differs by firm size. SMEs offer the largest AI premia for remote work and parental leave, along with positive premia for inclusive culture and health and well-being, but not for tuition assistance or paid leave. In S&P 500 firms, AI postings stand out only for inclusive culture language. Inclusive culture language is the one benefit where AI postings carry a premium in all three groups. Third, we find that the relationship between wages and benefits has evolved substantially over time. Prior to the emergence of generative AI, complementarity between high wages and non-monetary benefits was largely confined to resource-rich S&P 500 firms, whereas SMEs more frequently relied on benefits as partial substitutes for wages. Following 2022, however, the differences across firm sizes shrink, and complementary compensation strategies become increasingly widespread across firms of all sizes.

These findings suggest that as AI demand rises, there is not only a growing prevalence of benefit offerings but also changing compensation dynamics across firms. As AI capabilities become valuable throughout the economy, employers increasingly offer multidimensional compensation packages rather than salaries alone. Despite the growing importance of AI in the labor market, little empirical work has examined how firms combine monetary and non-monetary rewards to recruit AI talent, or how these strategies differ by firm size. Research has mostly examined pay and benefits separately, and leaves open why firms settle on a particular mix of rewards and the impacts (Fulmer & Li, 2022). Our study helps fill this gap and contributes to the literature on AI skill premia, employer competition, labor market signaling, and workforce transformation. Taken together, our study provides evidence on how firms are reshaping the terms of compensation for skilled labor alongside the diffusion of generative AI. We show that these changes extend beyond wages to the non-monetary dimensions of job quality; that small and medium-sized firms, not only frontier technology companies, are central to these shifts; and that the combination of high pay and generous amenities once reserved for the largest employers is spreading across the firm-size distribution. To our knowledge, ours is the first large-scale analysis of a range of non-monetary benefits advertised in AI job postings. These patterns have implications for job quality, for the inclusiveness of the growing AI workforce, and for policy on flexible and family-friendly work arrangements. The remainder of the paper is structured as follows. Section 2 reviews the relevant literature. Section 3 describes the data and methodological approach. Section 4 presents the empirical findings, and the final sections discuss implications for firms, policymakers, and future research.

## 2. Background

### *2.1. Finding AI talent is becoming a challenge*

Demand for AI skills has grown rapidly and spread well beyond the technology sector. AI adoption among firms has accelerated rapidly, rising from around 7% to 20% across OECD countries between 2021 and 2025, with higher uptake among larger and more innovative firms (OECD, 2026). Alekseeva et al., (2021) show AI-related job demand rose across most industries and occupations, with the highest intensity in IT, engineering, science, and management roles. Acemoglu et al. (2022) document a similar expansion using Lightcast data as used in this study, and show the growth in AI roles in firms exposed to AI is accompanied by decreased demand for non-AI roles. Squicciarini & Nachtigall (2021) document similar growth across Canada, Singapore, the United Kingdom, and the United States and across sectors, and find AI roles increasingly require diverse competencies like communication, problem-solving and teamwork. Recent work suggests that the digital economy and technological innovation continue to generate demand for new digital capabilities and raise employment overall, despite concerns regarding technological displacement (Uddin et al., 2026).

Meanwhile, finding and retaining talent is a major challenge for firms. On the supply side, uncertainty in hiring arises from difficulties in predicting future skill needs and turnover, combined with little control over future talent availability (Cappelli, 2009). Many companies have relied on "just-in-time" external hiring instead of internal development or workforce planning, but this strategy can lead to talent gaps when labor markets tighten (Cappelli & Keller, 2014). Employers must adjust what they ask for based on demand, and raise skill requirements when workers are plentiful (Modestino et al., 2020). At the same time, educational institutions struggle to adapt quickly enough to equip new talent with needed skills (Collins & Halverson, 2018; Stephany, 2021), and WEF 2025 reports that 63% of employers identify skills gaps as the largest obstacle to transforming their businesses (World Economic Forum, 2025).

Generative AI is also reshaping the nature of work, which the labor market must react to. Bone et al. (2025) emphasize that AI roles often require a broader skill set than other jobs. In the U.S., complementary socio-emotional, leadership, and problem-solving skills are especially valuable alongside technical AI skills (Borgonovi et al., 2023). As AI capabilities diffuse across industries, workers possessing hybrid technical and interpersonal competencies become increasingly valuable (Chang & Seo, 2026). LLMs have the potential to impact a large share of tasks in jobs, meaning they can have an important impact on the labor market (Eloundou et al., 2024). Devigili et al. (2026) show that the type of skills demanded is shifting based on the capabilities generative AI can offer, as organizations reorganize and reduce demand for skills the technology can perform. Firms must therefore also find talent suited to the new tasks, increasing demand for newly prioritized skills. Firms are also betting on productivity gains from AI, with evidence for the tools' improvement on productivity including larger gains for workers with less experience (Brynjolfsson et al., 2025). Jensen and Schneider (2026) find that firms seeking a larger share of digital skills in their vacancies are more productive, especially when those skills are in data science and AI, and that this demand is followed by growth in total employment. This can motivate firms to hire for AI skills, which can improve their productivity and bottom line. Despite rapid AI adoption, workers with specialized AI skills remain rare, accounting for only around 1% of the workforce, while shortages of relevant skills remain an important constraint on firms' ability to adopt AI technologies (OECD, 2026).

Empirical evidence underscores the rewards for AI talent. Alekseeva et al. (2021) find significant wage premia for AI skills compared to non-AI roles, both within the same firm and across the same job title. The demand for AI skills further intensifies for firms with more resources. Bone et al. (2025) also found that AI skills had a roughly 23% wage premium — a higher premium than having a master's degree — while Garcia-Lazaro et al. (2025) showed a 5.8% premium for those with digital skills in the same role, and around a 9% premium for listed AI and cybersecurity skills in the UK. In short, at this pivotal moment in AI's development, the talent shortage is acute: the skills gap is widening, demand for AI talent is extremely high, and monetary rewards for AI skills have risen.

### *2.2. Benefits are part of a total compensation package*

This section draws on three perspectives to explain why firms advertise non-monetary benefits and how this may differ across firms. Compensating-differentials theory treats a job as a bundle of pay and non-wage attributes, in which better amenities can offset lower wages (Rosen, 1986). Signaling theory explains how job seekers infer what an employer is like from the information firms send in job advertisements (Rynes et al., 1991; Uggerslev et al., 2012). Institutional theory explains why some practices become standard across a field and therefore signal little about a particular employer (DiMaggio & Powell, 1983). Together they suggest that the benefits a firm advertises depend on what workers value, what the firm can afford, and what the benefit communicates beyond what other employers already offer.

### *2.3. Workers value job characteristics beyond wages*

Monetary rewards alone are not the only means to attract skilled employees and address talent shortages. Workers also value features related to job quality such as flexibility, leave provisions, working conditions and organizational climate, and they evaluate jobs as a bundle of both monetary and non-monetary attributes (Mas, 2025). Non-wage compensation is therefore broader than conventional benefits. Beyond the core benefits of health cover, pensions and paid time off, it covers how schedules are set, family leave, the option to work from home and the physical work environment (Mas, 2025).

Non-monetary benefits are also an important part of compensation and firm expenses: in 2020, employer-paid benefits for the median U.S. worker were worth close to half of wages (Lavetti, 2023; U.S. Bureau of Labor Statistics, 2020). Employers and employees increasingly treat pay and benefits as a single package, which recent reviews organise around pay, retirement and health benefits, and benefits that support work-life balance (Fulmer & Li, 2022; Gandasari et al., 2024; Zingheim & Schuster, 2000).

Maestas et al. (2023) found that workers were substantially willing to trade pay for job amenities, though there is variation across worker demographics; for instance, workers earning higher wages place greater value on flexibility and have generally better working conditions and amenities. Preferences can also differ by gender; in Wiswall and Zafar's (2018) choice experiment, women gave up more pay for flexibility and job stability, while men paid more for faster earnings growth. Experimental evidence shows that average workers would take a pay cut in order to have remote work (Mas & Pallais, 2017).

More recent work has expanded this perspective by demonstrating that flexibility, autonomy, career development, remote work, organizational culture, and fringe benefits constitute economically important components of total compensation that shape workers' job choices and reservation wages (Lamadon et al., 2024; Mas, 2025; Mas & Pallais, 2017). Consequently, employers offer multidimensional compensation bundles rather than salary alone.

### *2.4. Job and organizational attributes attract applicants*

Providing desirable benefits makes a company more appealing to job seekers. Perceived fit has been found to be the strongest predictor for applicant attraction, followed by job characteristics, including total compensation with pay and benefits, and organizational characteristics (Uggerslev et al., 2012). Perceived benefits had a positive, significant impact on applicants' intention to apply for a role (Gandasari et al., 2024), and job advertisements with flexibility on when and where to work attracted more applications at the same salary (He et al., 2021). A strong employer brand also positively influences application intentions (Barber, 1998; Gandasari et al., 2024).

Benefits also matter after hiring. Employee benefits were positively associated with job satisfaction in the hospitality industry (Yoopetch et al., 2021), and well-being initiatives and non-monetary incentives can reduce turnover by improving employees' happiness and commitment (Rodríguez-Sánchez et al., 2020). Offering more vacation days was also found to be associated with lower voluntary turnover in Canadian tech firms (Renaud et al., 2021). Applicants respond most strongly when salary is paired with softer job features such as advancement prospects and a good work climate (Baum & Kabst, 2013).

Offering remote work flexibility also enables firms to attract more scarce talent from a larger pool (Van Landeghem et al., 2024). Remote work may also broaden firms' recruitment options when skills are scarce; qualitative evidence suggests that some firms adopt remote-working arrangements in part to address human-resource and skills shortages (Soroui, 2021).

### *2.5. Benefits in job postings can act as recruitment signals*

From a theoretical perspective, explicitly offering attractive benefits in job postings can be viewed through the lens of signaling theory. Because most US benefits are voluntary rather than legally required, the package a firm offers, and how much of its cost it passes on to employees, is largely its own decision (Fulmer & Li, 2022). Because benefit provision is largely discretionary, packages vary across employers, and the benefits a firm chooses to advertise reflect a strategic decision about how to recruit workers.

In recruitment, job seekers have limited information about employers, so they interpret the signals companies send – including details in job advertisements – to infer what working there might be like (Rynes et al., 1991; Uggerslev et al., 2012). Organizations therefore strive to send appealing signals early in the hiring process to attract suitable candidates (Backhaus, 2004; Schuth et al., 2018). Applicants can interpret signals, which affect organizational attractiveness, as cues to the pride they would get from working there, alignment with their values, and how they believe they would be treated as employees (Jones et al., 2014). For instance, listing flexible work options or wellness programs in a job ad signals that the company values employee well-being, which may attract candidates who share those values.

The person–organization fit literature suggests individuals are drawn to employers whose values and culture align with their own, and companies often use benefit and culture statements in job descriptions as signals to entice applicants and convey alignment (Backhaus, 2004). Workers will accept lower wages in exchange for fringe benefits, but differ in how much they value them and sort into jobs according to the benefits they value most (Eriksson & Kristensen, 2014), so employers can design reward packages aimed at the workers they most want to hire and retain (Fulmer & Li, 2022). Rich information about benefits and intangibles in postings can help candidates self-select into organizations where they will thrive, and helps firms differentiate themselves in a competitive market.

### *2.6. A new generation of benefits emerges*

Traditional benefits packages were largely designed for baby boomers and emphasized health insurance and retirement plans, whereas younger cohorts place higher value on different perks and flexibility (Clark, 2007) and value leisure more than earlier cohorts did (Twenge et al., 2010). Both men and women, especially younger professionals, highly value work-family balance and flexibility, and organizations trying to attract more diverse tech talent have found that emphasizing work-life balance is effective for drawing high-skilled women and men alike (Schuth et al., 2018). Because each generation forms its expectations of employers under different labour-market conditions, a single benefits package is unlikely to appeal equally to all workers, and firms may gain more by tailoring benefits to the groups they most want to attract (Dencker et al., 2007, as cited in Dulebohn et al., 2009).

Non-monetary rewards have become a growing part of pay packages in many countries (Eriksson & Kristensen, 2014), and employer spending on benefits in the United States has risen sharply over recent decades (Lavetti, 2023). In the early 1990s, few U.S. establishments offered work–family programs other than flexible hours (Osterman, 1995). On top of the standard package of pay, paid time off, health insurance and retirement plans, firms increasingly use discretionary benefits, from telecommuting and care subsidies to wellness programmes and student loan repayment, to target the workers they want to recruit and keep (Fulmer & Li, 2022). Rodríguez-Sánchez et al. (2020) argue that modern employees value benefits and incentives that enhance their well-being and job satisfaction, reflecting broader societal changes. Employers have meanwhile passed an increasing share of health insurance and retirement costs on to employees (Dulebohn et al., 2009; Fulmer & Li, 2022), so newer discretionary benefits sit alongside traditional ones that have become less generous. After health and retirement benefits, leave benefits, flexible working arrangements, family-friendly benefits and professional development benefits have been ranked among the most important benefits by employers (SHRM, 2024).

### *2.7. Wages and benefits: substitutes or complements*

Under compensating-differentials theory, workers accept lower wages in exchange for desirable non-wage job characteristics, treating the two as substitutes (Rosen, 1986). Search models argue that when workers can change jobs, firms compete through combinations of monetary compensation and non-monetary job amenities, with workers valuing the overall utility of employment beyond only wages (Hwang et al., 1998).

Consistent with this, Baum and Kabst (2013) found an inverse relationship between salary and soft factors, suggesting that higher pay often compensates for a lack of benefits, though salary alone may not fully substitute for them. Firms may likewise use corporate social responsibility, remote work and flexible hours to attract needed talent when job seekers would otherwise require substantially higher wages (Van Landeghem et al., 2024).

The relationship between wages and non-monetary benefits, however, can depend on firms' resource endowments. While the classical theory of compensating wage differentials predicts that attractive job amenities substitute for wages, recent evidence suggests that labour-market frictions and firm heterogeneity often generate complementarity instead. Firms differ in productivity, and with higher productivity, firms can afford to pay more and also provide better amenities, with the ability to spread the fixed costs of amenities across more workers (Mas, 2025). In U.S. data, firms with higher pay were shown to provide more amenities (Sockin, 2021). In Danish survey data, workers at higher-paying firms reported better perks and more flexibility. Their jobs also carried more layoff risk, a faster pace and more stress, which left the overall non-pay package slightly less valuable (Humlum et al., 2025).

By contrast, small and medium-sized enterprises (SMEs) typically operate under tighter financial constraints and more limited organizational capacity, including HR resources to invest in their workforce development (OECD, 2021). They tend to manage people more informally than large firms, with fewer established HR procedures (Cardon & Stevens, 2004; Kotey & Slade, 2005) and could respond to financial constraints by emphasizing organizational characteristics like flat structures or empowered ways of working that are comparatively inexpensive (Cardon & Stevens, 2004; Kotey & Slade, 2005; Way, 2002). Such non-monetary rewards can enhance the attractiveness of employment offers without requiring the substantial and recurring financial commitments associated with permanent wage increases or costly benefits such as paid parental leave or tuition assistance. Moreover, highly skilled workers increasingly value intrinsic aspects of work — including autonomy, meaningful work, and opportunities for professional growth — alongside monetary compensation (Deci & Ryan, 2000; Parker et al., 2017).

The emergence of generative AI may amplify these differences. Firms with fewer employees face gaps in skills, funding and digital infrastructure that slow AI adoption, and those with 10-49 employees use AI at under a third of the rate of large firms (OECD, 2025). As AI skills become valuable across a broader range of firms, the search for scarce AI talent extends beyond frontier firms to resource-constrained

SMEs. We therefore expect firm size to moderate the composition of compensation packages offered for AI-skilled positions. While resource-rich firms are likely to complement wage premia with extensive non-monetary benefits, SMEs are expected to rely relatively more heavily on non-monetary amenities as a cost-effective mechanism for attracting and retaining AI-skilled workers. This suggests that the relationship between wages and non-monetary benefits is shaped not only by the scarcity of AI skills but also by firms' financial and organizational resources.

### *2.8. Six non-monetary benefits*

In this study, we focus on six categories of non-monetary benefits: health and well-being, parental leave, tuition assistance, paid leave, inclusive culture, and remote work. These categories were selected based on their theoretical relevance and practical salience in current labor market research. Each of these benefits represents an important aspect of the broader “total rewards” framework (Zingheim & Schuster, 2000), encompassing both tangible and intangible factors that contribute to employee attraction and retention.

Paid leave continues to be a foundational employment benefit, often associated with reduced turnover and improved job satisfaction (Renaud et al., 2021). Tuition assistance reflects investments in employee development and has been linked to increased employee loyalty and performance (Barber, 1998). In one large employer, it increased retention by changing who joins the firm (Manchester, 2012). Prior studies have emphasized the role of health and well-being programs in enhancing job satisfaction and reducing burnout (Pfeffer et al., 2020). Inclusive culture, though more abstract, plays a crucial role in signaling organizational values and in shaping perceptions of person-organization fit, a well-documented determinant of applicant attraction (Backhaus, 2004). Parental leave and workplace flexibility, including remote work, have been identified as critical for work-life balance, especially for employees with caregiving responsibilities (Rodríguez-Sánchez et al., 2020; Schuth et al., 2018).

The six benefits differ in how much they cost employers and whether they commit the firm to a particular provision. Tuition assistance, paid leave, and parental leave entitle employees to specific compensation and involve dedicated planning and administrative infrastructure, whose per-employee cost decreases with economies of scale. Well-being benefits can vary widely on cost, but some employers have scaled back costlier wellness provision in favor of cheaper, occasional offerings such as meditation or stress-management sessions and on-site massages (Lester et al., 2021). We intentionally excluded traditional health benefits, such as health insurance, which is costly and can be legally mandated, to study more modern, optional well-being offerings such as well-being stipends and mental health support.

When it comes to remote work, the organizational cost to employers can vary. In the U.S., tech workers would take a 25% pay cut on average for a job offering remote or hybrid work modalities, yet companies often provided comparable salaries for the same role whether it was in-person or remote (Cullen et al., 2025). Employers and workers can be misaligned on the value of these arrangements. Lewandowski et al. (2026) show that in Poland, many employers perceive the costs of remote work above workers' perceived value of the benefit. Nonetheless, 25-35% of managers considered remote work to provide at least as much productivity as in-office work, and among these there was a smaller difference between perceived costs and employees' perceived benefits. While some firms may perceive costs in productivity or digital infrastructure, the benefit can be offered without expensive planning or administrative capacity that other entitlements require, particularly for small firms who can save on office costs.

For inclusive workplaces, the channels through which this is realized can vary, and it differs from the others as naming it doesn't commit the firm to a specific provision. It functions mainly as a signal of organizational values, and naming it in a job description itself is less costly for employers. We therefore treat tuition assistance, paid leave, and parental leave as costly entitlements, health and well-being and remote work as variable-cost benefits, and inclusive culture as a less costly signal.

These benefits also differ in the kind of information they carry. Institutional theory explains how organizational practices become standard across a field: firms converge on the same practices through legal and political requirements, imitation of other firms under uncertainty, and professional norms (DiMaggio & Powell, 1983). Offering benefits that have converged in this way therefore provides little competitive advantage or signal from an employer. Health insurance and retirement plans are among the most widespread and costly employer benefits (Dulebohn et al., 2009), whereas work–family benefits such as flexible schedules and childcare support have historically been offered by far fewer employers (Osterman, 1995, as cited in Dulebohn et al., 2009). Paid leave is highly valued by workers (Maestas et al., 2023) but widely provided, which limits what it signals about a particular employer. In 2026, 77% of U.S. civilian workers had access to paid vacation and 83% to paid sick leave (U.S. Bureau of Labor Statistics, 2026). Tuition assistance is a common benefit, historically offered by close to half of firms employing over 20 people (Black & Lynch, 1998, as cited in Manchester, 2012). Well-being benefits, parental leave, remote work, and inclusive workplace therefore function as differentiating benefits with meaningful signal from employers.

We intentionally include this range of benefits to distinguish whether firms advertise to talent via costly offerings or inexpensive signals, and test how this varies with firm resources. While we initially considered additional categories such as in-house childcare, career development programs, and travel assistance, these benefits were excluded from the final analysis due to limitations in the reliability and consistency of identification within the job vacancy data. By focusing on the six benefit types listed above, we ensure both conceptual relevance and empirical reliability in our analysis of how firms use non-monetary incentives to recruit for AI talent.

The literature clearly establishes that non-monetary benefits can serve as powerful tools to attract, motivate, and retain employees. However, most prior studies have examined benefits using surveys or theoretical arguments. To our knowledge, no study has analyzed the prevalence of diverse non-monetary benefits explicitly offered in online job vacancies, nor compared these offerings for roles requiring in-demand skills (like AI skills) versus other roles. This paper contributes to filling that gap by providing an innovative analysis of multiple non-monetary benefits offered in online job descriptions for AI-related roles, using a large dataset of job postings. By examining actual job ads, we gain direct insight into how organizations signal their value propositions to coveted AI talent in the competitive labor market.

### *2.9. Research question and hypotheses*

In light of the rising demand for AI talent and the growing relevance of non-monetary benefits in recruitment, this paper asks: **Do firms offer non-monetary benefits to attract AI talent?** In other words, beyond offering higher salaries for AI roles, are companies also highlighting additional perks and benefits in those job postings?

To address this question, we examine differences in advertised benefits between AI and non-AI job postings. We propose the following hypotheses:

- H1a: *AI-related roles are more likely to include non-monetary rewards in their postings than other roles, regardless of other relevant job characteristics.*
- *H1b: We assume this effect to be weaker for non-monetary rewards that impose significant and immediate costs, such as tuition assistance, paid leave, and parental leave.*

We also expect the combination of monetary and non-monetary premia to AI skills to vary with firms' organizational resource endowments. We present two further hypotheses:

- H2a: *In capacity-constrained firms (e.g., SMEs), non-monetary rewards are more likely to substitute for wage premia in the compensation of AI-related job roles.*
- H2b: *In capacity-rich firms (e.g., S&P500), non-monetary rewards are more likely to complement wage premia in the compensation of AI-related job roles.*

In the next section, we describe our methodology for analyzing job posting data and identifying AI roles and benefits to test these hypotheses. Section four presents the results of our investigation and the final sections discuss the results, implications, and conclusions.

## 3. Data and methods

To test our hypotheses, we use a dataset of online job vacancies, described in detail below. We first summarize the structure and limitations of the data, then explain our skills-based approach and how we identified AI-related roles and benefits in the dataset.

### *3.1. Online job vacancies*

Our primary data source is the Online Job Vacancy (OJV) database compiled by Lightcast (formerly Burning Glass Technologies) and made available via the Burning Glass Institute, a non-profit labor market research organization. Lightcast's OJV data has been widely used in empirical labor market research to study the rising demand for AI skills (Acemoglu et al., 2022; Alekseeva et al., 2021). The data is aggregated from approximately 65,000 sources, including company websites, job boards, and job aggregators. Vacancies are then deduplicated, cleaned, and standardized into an analyzable format (Lightcast Press Office, 2023). We use online job postings as they capture the offers firms advertise to candidates, and they are available at a scale that can capture labor market variations at a magnitude that surveys do not match while capturing a variety of variables over time.

We pull a random sample from 2018 to 2025 of ten million job postings (N=9,999,673 after dropping blank rows) from the U.S., associated with approximately 137 million listed skills (137,486,868), 28,939 of them unique. Each listing includes structured metadata such as job title, location, employer name, and industry. Crucially, the postings also provide granular detail on employer preferences, including required skills and education levels. For a subset of postings in the United States, advertised salary information is available for approximately 36% of the sample. Salary data is distributed similarly for roles that require AI skills (35.2%) and those that do not (35.6%).

While this dataset provides extensive insights into employer demand, it has limitations. First, we observe only the job postings — not the selection or hiring process — so unlisted selection criteria may influence hiring decisions. Second, our analysis is limited to online postings and does not capture alternative hiring strategies, such as internal upskilling or reliance on external contractors. Third, online job ads are not uniformly distributed across the labor market; some occupations and industries are more heavily represented than others.

### *3.2. A skill-based approach to identify AI roles*

To identify AI-related jobs, we adopt a bottom-up, skill-based classification strategy. Unlike top-down methods that define occupations based on sectoral affiliation (e.g., classifying anyone at a tech company as an AI worker), bottom-up approaches define roles by the actual skills or tasks required. In our context, a job is considered AI-related if the posting explicitly mentions at least one AI skill — regardless of the company's industry. We test the robustness of this approach by reviewing results with additional thresholds of two and three AI skills — as shown in Figure A5 — and find that the direction and general magnitude of coefficients hold.

The use of task- or skill-based approaches aligns with broader trends in labor market research. Studies dating back to the 1990s introduced the concept of Skill-Biased Technological Change to explain rising wage inequality (Goldin & Katz, 2007; Katz & Murphy, 1992). Later research emphasized the growing importance of high-skilled workers for leveraging new technologies, particularly in computer-augmented roles. This line of inquiry eventually led to more nuanced models that examined how evolving skill requirements impacted different segments of the labor force (Autor et al., 2003; Manning, 2004).

Acemoglu and Autor (2011) proposed a task-based framework to better capture how technological advances reshape labor demand and wage structures. This approach laid the foundation for recent research that uses job-level data to analyze the relationship between technology and skill requirements at a granular level (Acemoglu & Restrepo, 2019; Alekseeva et al., 2021; Stephany & Teutloff, 2024). Following this tradition, we use Lightcast's Open Skills Taxonomy, which classifies over 30,000 unique skills, to identify AI-related vacancies. Specifically, we rely on Lightcast's up-to-date "Artificial Intelligence and Machine Learning (AI/ML)" subcategory, which includes 279 distinct skills. Any job listing that mentions at least one of these skills is coded as an AI job.

This skill-based approach offers several advantages. It allows us to capture not only obviously AI-centric roles (e.g., machine learning engineers), but also positions in other domains (e.g., marketing, finance) that increasingly require AI capabilities. It also enables us to analyze and compare differences in salary, education requirements, and benefit offerings between AI and non-AI roles. Furthermore, this method is flexible and can be replicated in other countries and sectors, enhancing its utility for comparative studies.

### *3.3. A keyword approach to identify benefits*

To analyze non-monetary benefits, these had to be extracted from the online job vacancy descriptions. We chose a keyword approach after considering a variety of methods, including LLM-based classification, as it is an accessible, transparent, and auditable method that proved to achieve performance comparable to more costly alternatives.

We looked at health and well-being, parental leave, tuition assistance, paid leave, inclusive culture, and remote work as non-monetary benefits that may be offered by companies. For health and well-being, we specifically look at well-being initiatives that companies offer beyond health insurance, such as mental health benefits, well-being stipends, and gym memberships. This goes beyond the costly health insurance benefits that many firms must offer to examine more variable, values-driven well-being offerings. For parental leave, we look at maternity, paternity, and caregiver leave. For inclusive workplace cultures, we look for terms such as "inclusive environment" and "diversity and inclusion" advertised in the job description. Paid leave includes paid time off, mental health days, paid holidays, company holidays, and sick leave. Tuition assistance involves tuition assistance and reimbursement, and remote work includes both fully remote and hybrid work.

To ensure the accuracy of the classifications, we employed an iterative review process. An initial search of possible terms that could be used to indicate non-monetary benefits was conducted on the job descriptions of the job vacancies. Keywords were narrowed down to those that specifically identified benefits accurately and identified concrete, non-abstract offerings. We then manually reviewed a random sample of 100 jobs for each benefit, for a total of 600 jobs, to examine accuracy and identify any false classifications. Additionally, a set of exclusion keywords were added to exclude issue phrases that caused false positives. The final summary of keywords used is provided in the Appendix.

Meanwhile, this sample was checked for additional keywords for all benefits to identify any false negatives and ensure the list of keywords for all benefits was comprehensive. Subsequently, another random sample of 50 jobs per benefit was checked for accuracy. After this review process, we achieved over 95% accuracy for health and well-being benefits, parental leave, tuition assistance, paid leave, and inclusive culture. However, remote work posed an issue with false positives, particularly as after the pandemic, many job postings added a "remote work" field to indicate whether it was offered or not, thus generating many false positives where it was subsequently negated. An extensive list of keywords and exclusions was generated by iteratively checking the results to achieve higher accuracy. We then compared the results to a large language model that was developed to label remote work from online job vacancies (Hansen et al., 2023) and achieved a 94% agreement rate (F1 score of 0.805). We also checked all our analyses for robustness between the keyword method and the large language model and saw similar results. Thus, we chose to maintain the keyword method for consistency in methodology across benefits. Figure A1 in the Appendix illustrates the findings of the validation.

Following our initial keyword labeling, Lightcast added structured benefit labels to their data using a proprietary machine learning approach. For benefits from their taxonomy aligned with our benefit categorization — parental leave, paid leave, and tuition assistance — we subsequently checked performance against these labels. Each had an F1 score of around 94-95%, as shown in Table A4.

### 3.4. *Relating AI roles and benefits*

Our analysis proceeds in three stages to examine whether and how non-monetary benefits are used to attract AI talent in the labor market. In the first, descriptive stage, we assess the evolution of demand for AI roles relative to the overall labor market from 2018 to 2025. We identify the share of AI versus non-AI job postings that explicitly mention non-monetary benefits. We focus on six key benefit categories: tuition assistance, paid leave, health and well-being perks, parental leave, inclusive culture, and remote work options. For each benefit type, we compare the proportion of postings that offer the benefit across AI and non-AI roles and analyze how these ratios develop over time.

In the second stage, we conduct inferential analysis using logistic regression models to estimate the likelihood of a job posting including each specific benefit. The core explanatory variable is a binary indicator for AI roles. We run multiple model specifications. The baseline model includes only the AI indicator with industry and year fixed effects. The second model adds state fixed effects and controls for required education level and work experience. The third model includes posted salary to assess the independent effect of being an AI role on benefit offerings, and the fourth model adds an S&P 500 dummy to control for high firm productivity. A fifth model adds fixed effects for occupation codes. This approach allows us to assess whether the presence of benefits in AI postings is driven by the nature of the role or by confounding factors such as job level or compensation. We also examine the link between benefits and salary more directly by comparing the advertised wage levels in benefit-offering postings over time. The results of these analyses are presented in detail in the following section.

### *3.5. Distinguishing firms by size*

In the third stage, we further dissect the non-monetary offerings for AI roles by firm size to partially address heterogeneity by firm size and resources. We proxy firm size using the number of posts a firm has in our dataset in 2018. We only use data in 2018 to avoid endogeneity issues where firms that adopt AI over the period may grow faster and hire more relative to their firm size. Since this is an imperfect proxy, as firms can have different turnover rates for the same employment size, we validate our categorization approach using an LLM-as-a-judge. In particular, we decide on a post count threshold by sampling 1,000 employers from our dataset and ask GPT-6-luna with web search whether the firm is a SME (small or medium enterprise) or a large firm. The full prompt instructions used can be seen in the Appendix. Then we compare that against various thresholds and find that the post-weighted macro F1 score peaks at the threshold of 6. Therefore, we choose 6 posts as the cutoff threshold (for details see Figure A6 in the appendix). We keep S&P 500 firms as a separate category and exclude these firms from the large firm category to isolate the firms with the most resources.

# 4. Results

## *4.1. The demand and premium for AI roles are rising*

To represent AI demand in the OJV data, Figure 1 illustrates the evolution of employer demand for AI skills and the associated wage premium between 2018 and 2025. Prior to the introduction of generative AI in late 2022, AI skills were requested in approximately 1–1.5% of all job postings, while the estimated AI wage premium remained comparatively volatile. Following the release of ChatGPT in November 2022, both trends changed markedly: the share of AI-related vacancies increased rapidly to nearly 3.5% of all postings by the end of 2025, while the adjusted wage premium followed a clear and sustained upward trajectory, reaching almost 30%. The rise in both employer demand for AI skills and their labor-market value coincides with the emergence and rapid diffusion of generative AI. It also coincides with a broader transformation of the labor market post-pandemic, including increased digitalization and a reimagination of the workplace.

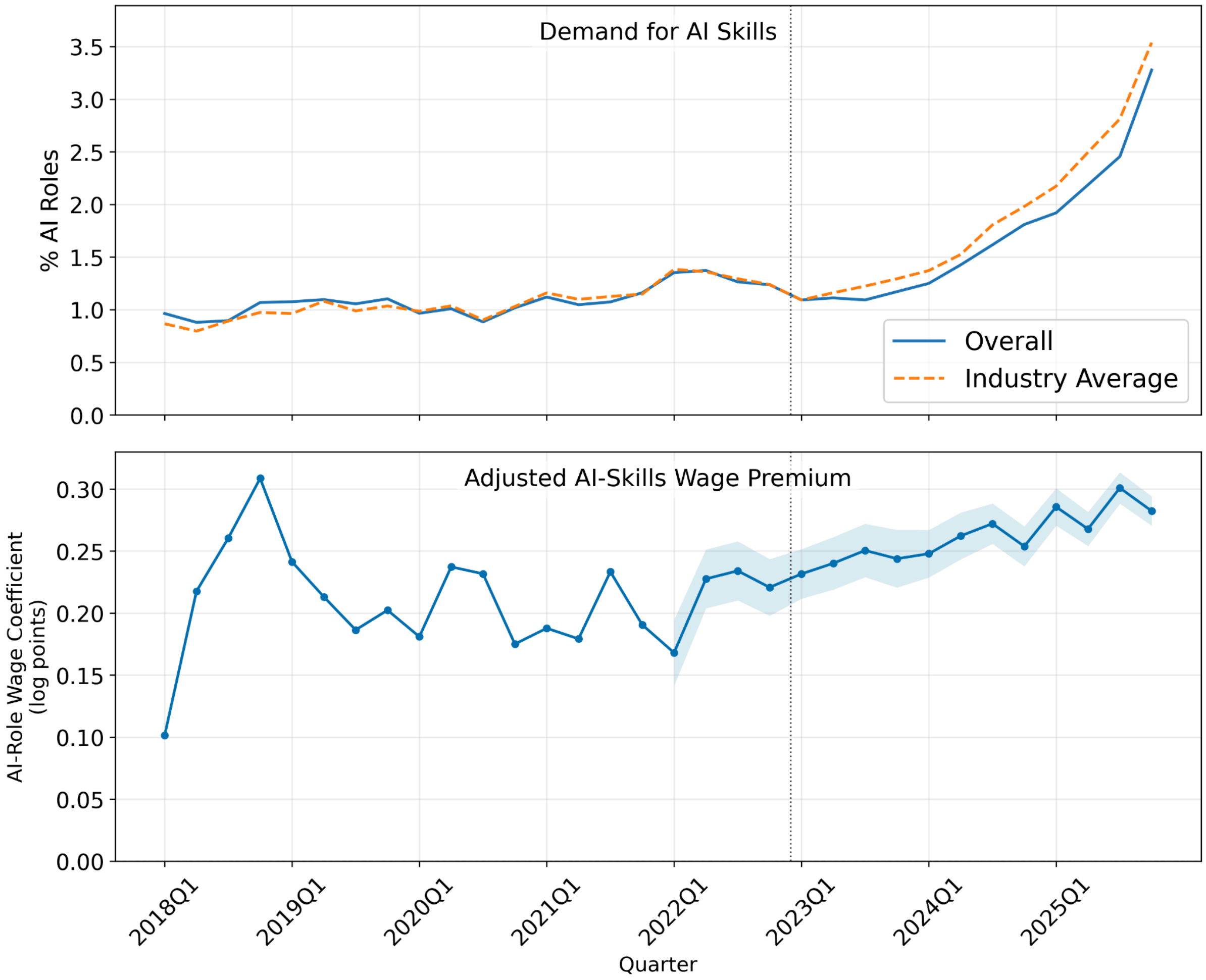

**Figure 1. Development of AI skill demand and wage premia in the U.S. labor market, 2018–2025.** The upper panel shows the share of online job vacancies requiring AI skills between 2018 and 2025. The vertical dashed line marks the release of ChatGPT in November 2022, after which demand for AI skills rises sharply from approximately 1% to nearly 3.5% of all vacancies by the end of 2025. The lower panel reports the adjusted wage premium associated with AI-skilled vacancies, estimated from regression models controlling for industry, education, experience requirements, state, and time fixed effects. While wage premia fluctuate considerably prior to late 2022, they exhibit a sustained upward trend thereafter, approaching 0.3 by the end of the observation period.

How does this continuous demand for AI roles relate to job benefits? Figure 2a (left) shows the percentage of jobs offering each benefit for both AI and non-AI roles. For nearly all benefits, AI roles have a higher proportion of jobs offering the benefit than non-AI roles. However, for tuition assistance and paid leave there is a slightly higher percentage of non-AI jobs offering these benefits. The difference between the role categories is largest for parental leave, inclusive workplace and remote work.

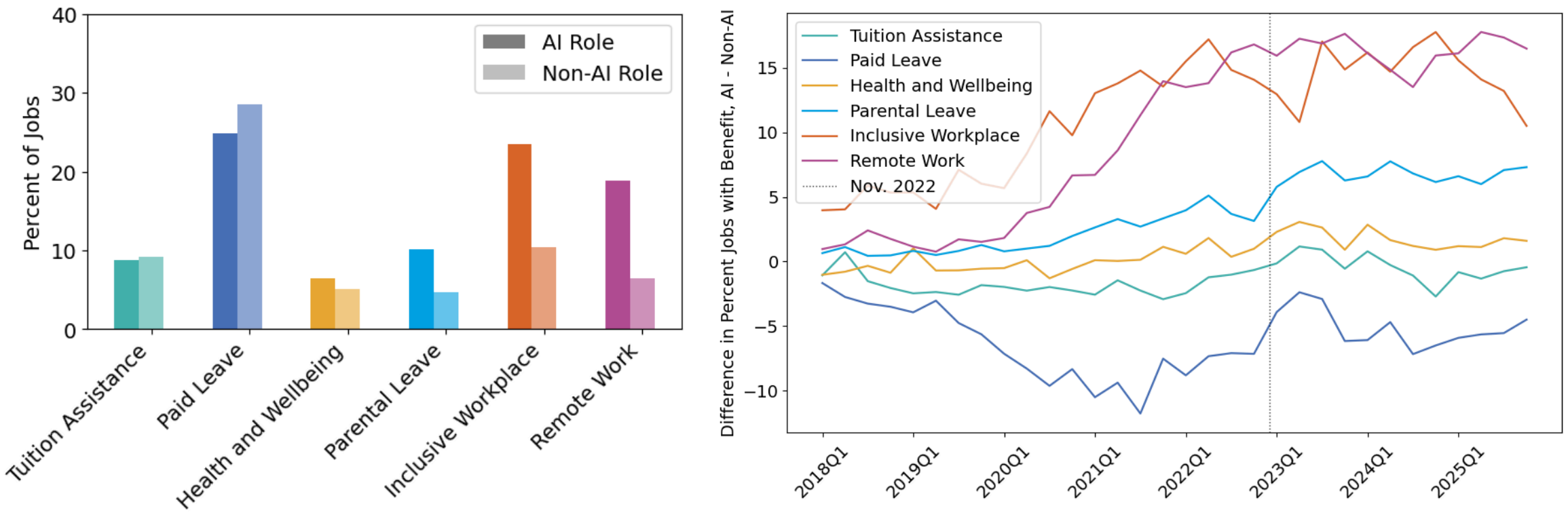

**Figure 2. Benefit Prevalence by Role Type and Temporal Change in AI vs. Non-AI Benefit Gaps. Left panel:** Share of job postings offering six key non-monetary benefits, separated by AI and non-AI roles. While some benefits — like paid leave — are common across both, others such as parental leave, inclusive culture, and remote work are significantly more likely to be offered in AI postings. **Right panel:** Temporal trend in the benefit gap, calculated as the difference in benefit prevalence between AI and non-AI roles. While the gap has declined slightly for paid leave and remained stable for tuition assistance and health and well-being, it has grown substantially overall for parental leave, inclusive workplace, and remote work, with variation between years.

Meanwhile, Figure 2b (right) shows the difference in the percent of jobs offering a given benefit for AI roles vs. non-AI roles over time. Inclusive workplace consistently shows a substantial difference between the role types, particularly since 2020, though it begins to decrease in recent quarters. Remote work increases after 2020 to similarly show one of the largest differences between AI and non-AI roles. Meanwhile, paid leave is consistently mentioned more often among non-AI roles than AI roles. The remaining three benefits — parental leave, health and well-being, and tuition assistance — have a much smaller difference between the two skill categories and generally remain within 5%, with parental leave increasing slightly in recent years.

### 4.2. *AI roles offer more benefits*

To investigate the role of job-level characteristics in shaping the likelihood of non-monetary benefits being offered, we estimate a series of logistic regression models, described by equation (1), where $\Lambda(\cdot)$ is the logistic CDF, $AI_i$ indicates an AI role, and $X_i$ is a vector of posting-level controls (indicators for required experience and education). δ, θ, and λ denote year, industry (3-digit NAICS), and state fixed effects.

$$\Pr(\text{Benefit}_i = 1) = \Lambda\big(\beta_0 + \beta_1\,\text{AI}_i + \beta_2\log(\text{Salary}_i) + \mathbf{X}_i'\gamma + \delta_{\text{Year}[i]} + \theta_{\text{Industry}[i]} + \lambda_{\text{State}[i]}\big) \quad (1)$$

The dependent variable in each regression is a binary indicator for whether the job posting mentions a specific benefit. We estimate the model separately for each of the six benefit types: health and well-being, parental leave, tuition assistance, paid leave, inclusive culture, and remote work. The key explanatory variable of interest is a binary indicator for whether the role is categorized as an AI role. In the baseline specification (Model 1), we include only industry and year fixed effects. In Model 2, we add controls for the required level of education (classified into five categories) and required work experience (classified into four levels). State fixed effects are included to control for regional variation, such as differences in regulation that can affect benefits. In Model 3, we additionally control for the log of salary offered in the posting, where available. Model 4 controls for firm size by introducing a dummy variable for S&P 500 firms. We tested adding firm fixed effects to further control for firm variability, but the models did not converge, likely due to the large number of employers combined with the sparsity of AI roles and benefit mentions within many firms. We therefore rely on industry fixed effects combined with an S&P 500 control to address some of this variability. Finally, we ran a fifth specification to add occupation fixed effects using SOC codes. This is an effort to address variation between occupations where benefits may be more common in particular occupations or be related to the type of work itself. This specification shows mostly similar results, showing robustness when accounting for this variability; only the remote premium decreases noticeably, but continues to exist. The full regression results for all six benefit categories are provided in Table A1 in the Appendix. As an additional robustness check, we compare the difference in the prevalence of benefits in AI and non-AI vacancies within each firm and occupation. Figures A11 and A12 indicate significant positive differences for AI roles compared to non-AI roles.

Figure 3 shows the coefficients for the AI role covariate for each of the five models across the six benefits. The y-axis shows the estimated log-odds coefficient along with 95% confidence intervals. Benefits like remote work, inclusive workplace, and parental leave have positive log-odds coefficients for AI-related roles across all models, suggesting that jobs in AI are significantly more likely to offer these benefits. Conversely, benefits like tuition assistance and paid leave have coefficients closer to zero or even negative, suggesting a weak negative association. As shown in the figure, the effect size of AI skills generally decreased with each model. This indicates that factors like education and experience partially account for the log odds, and part of the association may also be explained by salary. Nonetheless, parental leave, inclusive workplace, and remote work continue to have a positive, significant coefficient for AI skills, indicating partial support for hypothesis H1a: AI vacancies are significantly more likely to advertise non-monetary benefits. They also partially support H1b: the association is concentrated for remote work and inclusive workplaces — benefits requiring relatively fewer resources — and little or no association for the tuition assistance and paid leave. Parental leave is an exception, where we classify it as costly but it has a significant premium for AI roles, perhaps as a differentiating signal. Meanwhile, Panel B shows estimates for individual year models. Inclusive workplace, remote work, and parental leave remain positively associated throughout, with a large increase for remote work following 2020. Paid leave remains negative throughout, tuition assistance moves closer to zero, and health and well-being shifts to a positive association over time.

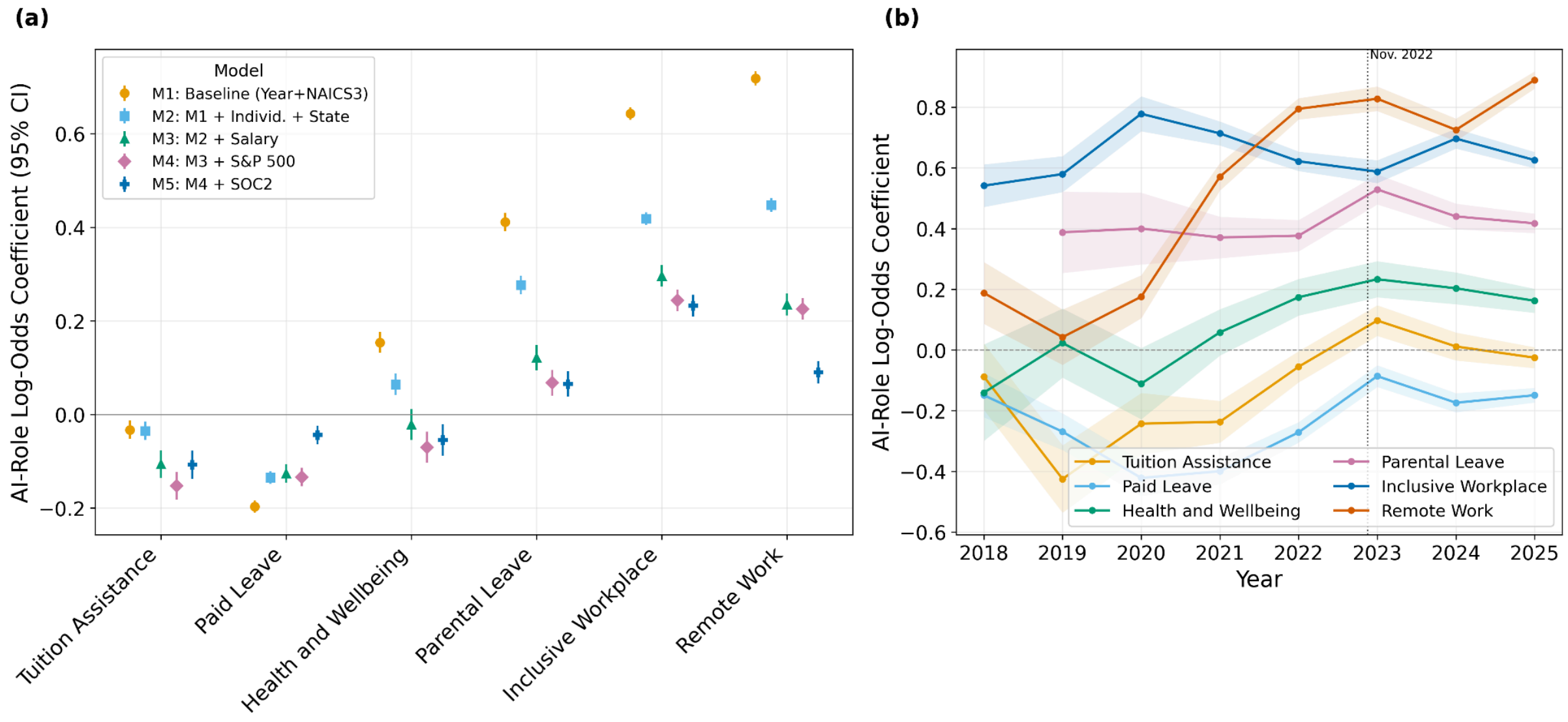

**Figure 3. Coefficients for Benefit Inclusion in AI vs. Non-AI Roles (Logistic Regression Models).** Panel A presents the coefficients from our logistic regression analysis (see Table A1), estimating the log

odds of observing each benefit in AI versus non-AI postings. AI roles are significantly more likely to include parental leave, inclusive workplaces, and remote work, even after sequentially controlling for job-level characteristics. While the effect sizes attenuate with additional controls, they remain statistically significant. Panel B shows the coefficients for M1 run on individual year models, dropping year fixed effects, showing the AI premium pattern over time.

### *4.3. The complementarity of benefits and wages*

To assess our second hypothesis — examining where non-monetary benefits are offered alongside, or instead of, higher salaries — we first examine the overall relationship between wages and benefit provision among AI and non-AI vacancies. We argue that in particularly tight labor markets, firms may enhance the attractiveness of positions by combining monetary compensation with non-monetary rewards. Here we investigate how this bundling of monetary and non-monetary benefits has evolved and if we can observe notable differences across firms of different sizes.

First, we examined whether benefits tend to appear together in a job posting. The bar chart in Figure A2 in the Appendix shows the number of jobs per the number of benefits offered. While the majority of jobs do not offer any benefits, those that do most often only offer one. Additionally, the correlation matrix shows the correlation between each benefit and salary on a job level. Salary information shows the highest correlation with paid leave. Paid leave has the highest correlation with tuition assistance benefits, followed by parental leave and health and well-being, showing that the most expensive, traditional benefits are paired together more often than with other low-cost benefits.

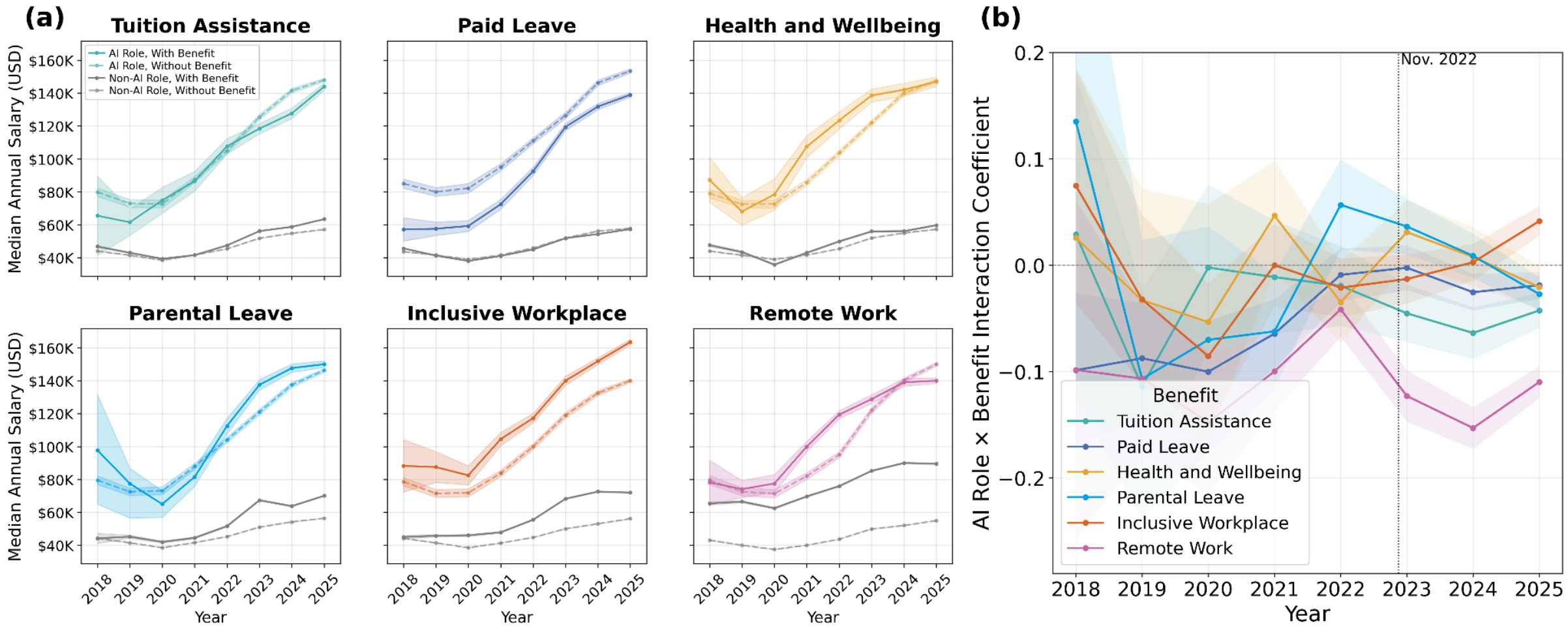

**Figure 4. The relationship between wages and non-monetary benefits in AI and non-AI vacancies, 2018–2025.** The left panel compares median offered annual salaries for AI and non-AI vacancies with and without six non-monetary benefits over time. The right panel reports interaction coefficients from wage regression models

estimating whether individual benefits complement (positive coefficients) or substitute (negative coefficients) the AI wage premium after controlling for industry, firm size, education, experience, geographic location, and time effects; error bars denote 95% confidence intervals.

Figure 4 summarizes this relationship using both descriptive and regression-based evidence. The left panel traces the evolution of median offered salaries between 2018 and 2025 separately for AI and non-AI vacancies, distinguishing postings that advertise a given benefit from those that do not. Consistent with our previous findings, wages associated with AI roles increase substantially over the observation period. Among non-AI vacancies, postings advertising parental leave, inclusive culture initiatives, and remote work consistently offer higher salaries than comparable vacancies without these benefits. For AI vacancies, however, the picture is considerably more nuanced. Benefits related to parental leave, inclusive culture, health and well-being, and remote work are generally associated with higher wages throughout much of the observation period, although these differences diminish or even reverse towards the end of the sample. In contrast, AI vacancies offering tuition assistance or paid leave frequently exhibit lower salaries than comparable AI vacancies without these benefits, suggesting that the relationship between wages and benefits differs substantially across benefit types.

To examine these patterns more rigorously, the right panel reports estimates from regression models explaining offered wages while controlling for industry, firm size, education requirements, seniority, geographic location, and time effects. For each benefit, we estimate the interaction between AI vacancies and the presence of the respective perk. Positive interaction coefficients indicate that benefits amplify the AI wage premium and therefore complement monetary compensation, whereas negative coefficients suggest that benefits partially substitute for higher wages.

The regression results broadly confirm nuances in the descriptive evidence while revealing an important shift over time. In recent years, the trend for most benefits has strengthened towards substitution. By the end of the sample period, only limited evidence remains that these benefits systematically enhance AI wage premia. The only benefit with increasing complementarity is an inclusive workplace.

Taken together, these findings suggest that the relationship between wages and non-monetary benefits is neither uniform nor static. Rather, employers strategically combine or substitute different forms of compensation depending on the type of benefit and prevailing labor market conditions. While AI roles continue to command substantial wage premia, the value of individual benefits relative to salary appears to evolve alongside the maturation of the AI labour market and the changing competitive environment.

### 4.4. *Benefit premia vary by firm size*

Having established that AI-related vacancies are associated with higher wages and more frequent non-monetary benefits, we next examine whether these relationships differ systematically across firms with varying resource endowments. Building on our theoretical argument, we distinguish between small and medium-sized enterprises (SMEs), large firms, and S&P 500 firms, allowing us to investigate how organizational resources shape compensation strategies before and after the emergence of generative AI.

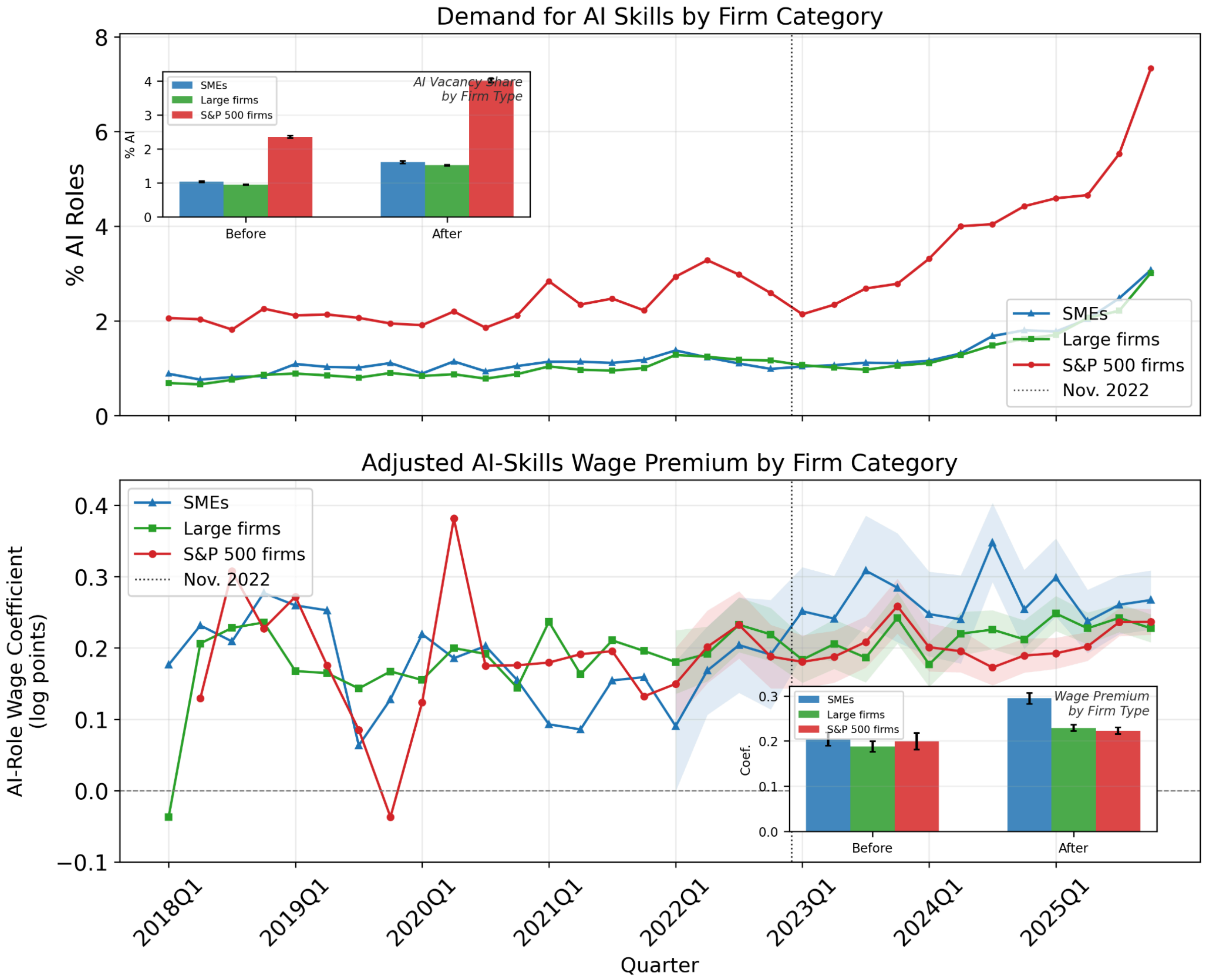

**Figure 5. Development of AI demand and wage premia across firm sizes.** The upper panel shows the share of job postings requiring AI skills among SMEs, large firms, and S&P 500 firms between 2018 and 2025. The lower panel reports the estimated wage premium associated with AI-skilled vacancies for each firm group. The vertical dashed line indicates the release of ChatGPT in November 2022, highlighting the acceleration in AI demand across all firms and the particularly strong increase in AI wage premia among SMEs.

Figure 5 first revisits the temporal evolution of AI labor demand and wage premia by firm size. The upper panel demonstrates that throughout the observation period, S&P 500 firms consistently exhibited the greatest demand for AI-skilled workers, with approximately 2–3% of vacancies requiring AI skills prior to the end of 2022, compared with roughly 1% among both large firms and SMEs. For both groups we register a significant uptake in AI roles following the introduction of generative AI. By the end of 2025, around 7% of vacancies posted by S&P 500 firms require AI skills, while demand among both large firms and SMEs has approximately tripled to around 3%. These findings suggest that AI capabilities shifted from a niche requirement concentrated among frontier firms into a broadly valuable input more often demanded throughout the economy.

Interestingly, the lower panel reveals a different pattern for wages. While AI wage premia increase across all firm types following the introduction of generative AI, the increase is most pronounced among SMEs. Prior to late 2022, AI skills were associated with wage premia of approximately 15-20% across all firm groups. Afterwards, SMEs exhibit a marked divergence, with the AI wage premium approaching 30%, exceeding the corresponding increases observed among larger firms.

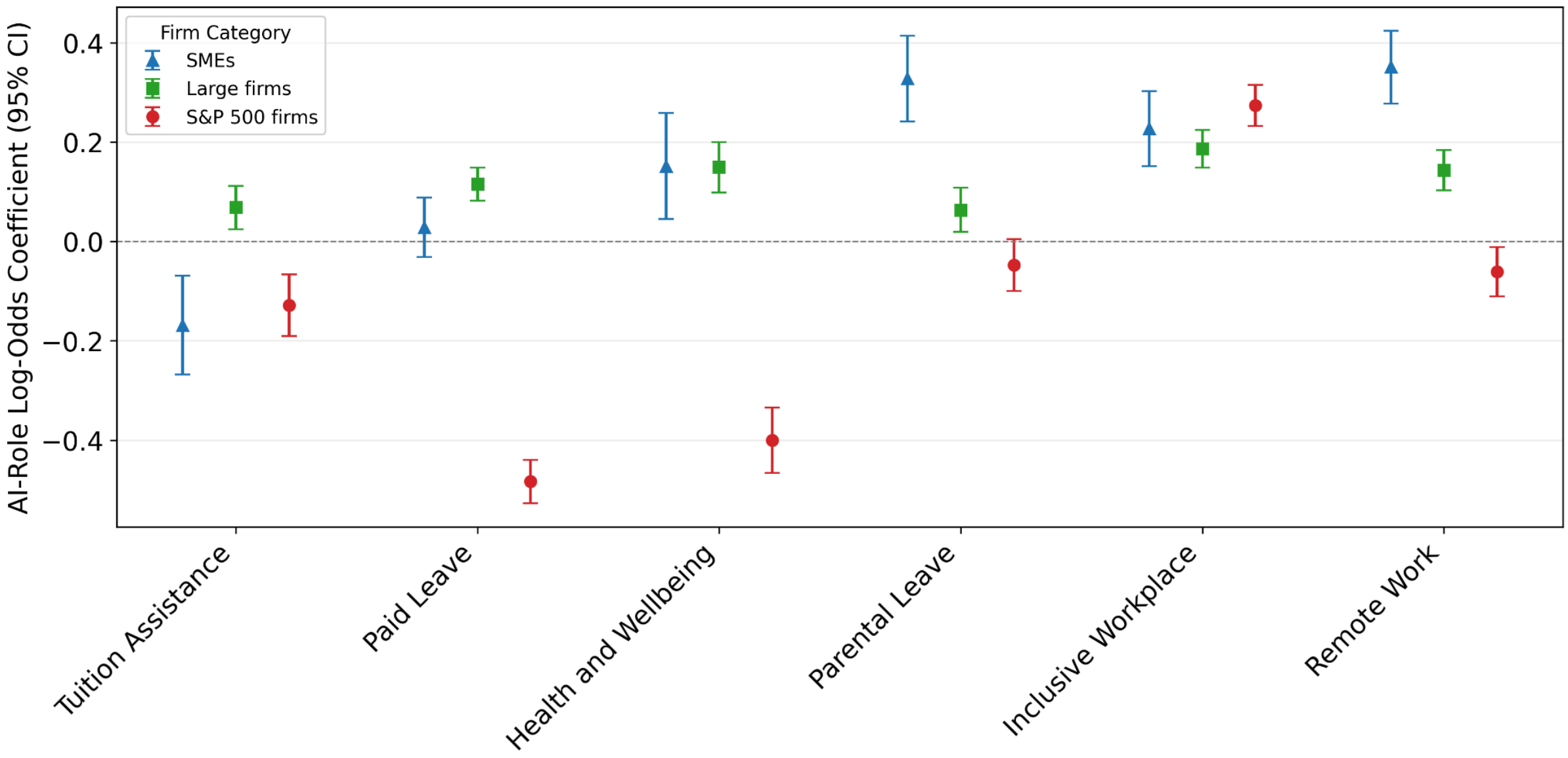

**Figure 6. Firm size heterogeneity in the provision of non-monetary benefits for AI-skilled vacancies.** The figure reports estimated coefficients from logistic regression models predicting the likelihood that AI-related vacancies offer specific non-monetary benefits, estimated separately for SMEs, large firms, and S&P 500 firms. Estimates are reported relative to comparable non-AI vacancies after controlling for industry, year, state, education, experience requirements, and salary (95% confidence intervals are shown).

We next investigate whether similar differences emerge in the provision of non-monetary benefits. Figure 6 shows that, when re-estimating our logit model separately for each firm category, SMEs offer a significantly positive premium for AI roles across remote work, parental leave, inclusive workplace, and health and well-being benefits, with the largest positive difference compared to other firm size groups in remote work and parental leave. By contrast, this difference disappears for more financially demanding and traditional benefits such as tuition assistance and paid leave, where large firms are the only group that offers a visible premium for AI roles. Large firms show a positive premium across all benefits, and as they contain 64% of the job postings, are a major driver of the overall premium. Meanwhile, S&P 500 firms only show a premium for inclusive workplace language.

These findings are partially consistent with our theoretical expectation that financially constrained firms emphasize non-monetary rewards with limited immediate costs. SMEs show no AI premium for tuition assistance or paid leave, but do show one for parental leave. They also strengthen the evidence that our initial findings, which show a benefit premium for AI roles, are not driven by a concentration of AI roles in the most productive firms.

Finally, we examine whether firms treat wages and non-monetary benefits as substitutes or complements. Following our baseline specification, we estimate wage regressions including indicators for AI vacancies, individual benefits, and their interaction, where positive interaction terms indicate complementarity between wage premia and benefits, while negative coefficients indicate substitution. Motivated by the pronounced structural break following the introduction of generative AI, we estimate the models separately for the periods before and after 2022, as shown in Figure 7.

$$\begin{aligned}\log(\text{Salary}_i) = \beta_0 + \beta_1\,\text{AI}_i + \beta_2\,\text{Benefit}_i + \beta_3\,(\text{AI}_i \times \text{Benefit}_i) \\ + \mathbf{X}_i'\gamma + \delta_{\text{Year}[i]} + \theta_{\text{Industry}[i]} + \lambda_{\text{State}[i]} + \varepsilon_i \end{aligned} \quad (2)$$

In our model, we explain $\text{Salary}_i$, which is the posted salary of vacancy i; $\text{AI}_i$ indicates an AI role, and $\text{Benefit}_i$ indicates that the posting offers the benefit in question. The coefficient of interest, $\beta_3$, captures the differential wage premium for AI roles that offer the benefit relative to those that do not. $X_i$ is a vector of posting-level controls comprising indicators for required experience and education. δ_Year[i], θ_Industry[i], and λ_State[i] are year, industry (3-digit NAICS), and state fixed effects. The model is estimated by OLS on the sample of postings with disclosed salaries, separately for each of the six benefits. We also run individual models for each year, omitting the year fixed effects, as seen in Figure A10.

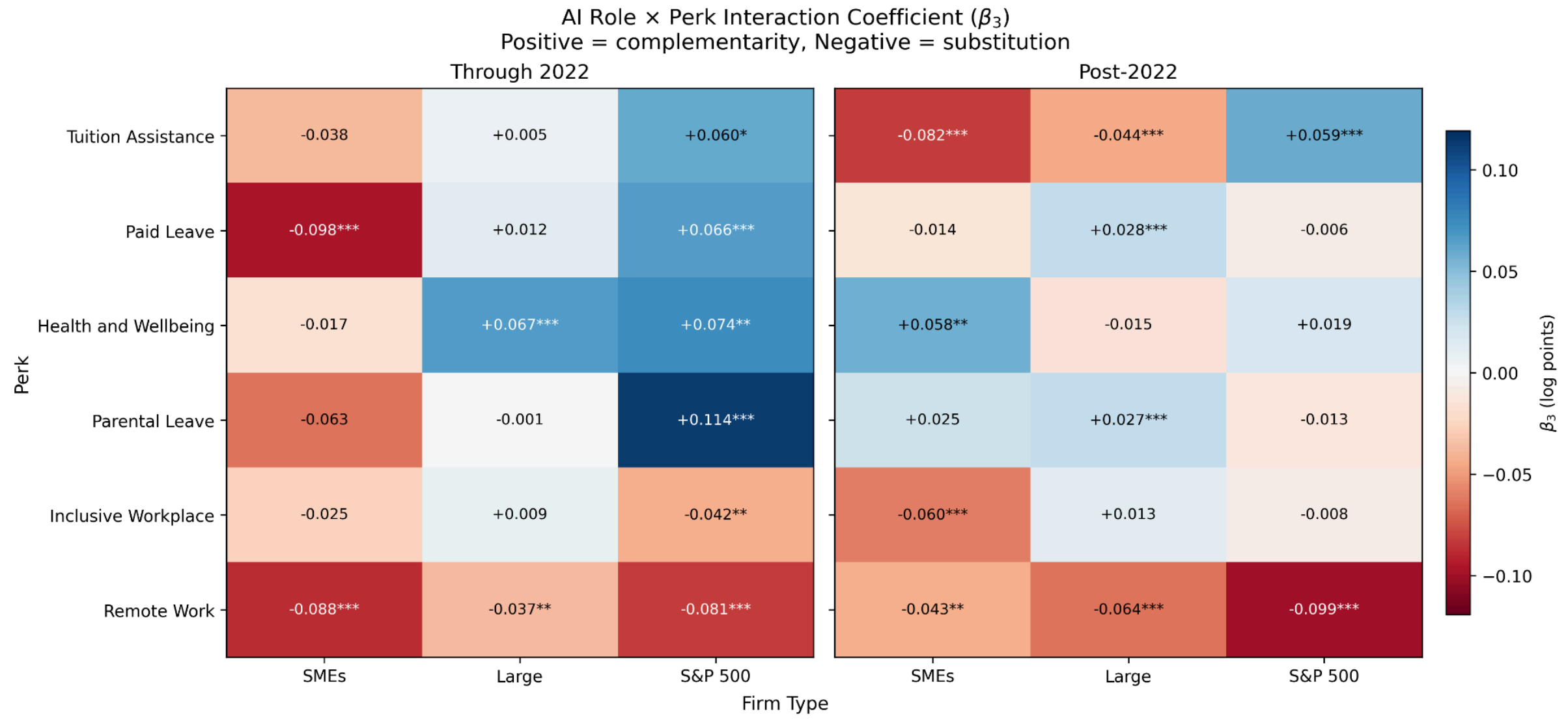

**Figure 7. Complementarity and substitution between monetary and non-monetary compensation for AI-skilled vacancies by firm size.** The figure presents interaction coefficients between AI-related vacancies and individual non-monetary benefits from wage regression models estimated separately for SMEs, large firms, and S&P 500 firms before (left panel) and after (right panel) the introduction of generative AI. Positive coefficients indicate that the presence of a benefit amplifies the AI wage premium (complementarity), whereas negative coefficients indicate that benefits partially substitute for higher wages. All models control for industry, education, experience, location, and year.

Before the emergence of generative AI, complementarity between monetary and non-monetary rewards was largely confined to S&P 500 firms. These firms simultaneously rewarded AI-skilled workers with higher salaries and costly benefits, including tuition assistance, paid leave, and parental leave, as well as health and well-being programs. In contrast, SMEs exhibited substitution effects for several of these benefits, suggesting that non-monetary rewards partly compensated for more limited wage increases. This pattern closely reflects the resource-based distinction proposed at the outset of the paper.

Following the introduction of generative AI, however, compensation strategies converge across firm types. Complementarity among S&P 500 firms mostly disappears, while large firms increasingly combine AI wage premia with paid leave and parental leave and reduce complementarity with health and well-being. Even SMEs begin to exhibit complementary relationships for selected benefits, most notably health and well-being, while the previously observed substitution effects for paid leave, parental leave, and remote work diminish or disappear. The most consistent substitution effect emerging across all firm types concerns remote work, which coincides with the stagnation of this benefit across roles as seen in Figure A3. H2a is therefore supported in the pre-2022 period, when SMEs exhibit clear substitution between wages and benefits while S&P 500 firms clearly complement. Post-2022, it is no longer supported, as SMEs and S&P 500 firms shift in opposite directions.

Whereas compensation strategies were previously sharply differentiated by firms' resource endowments, the rapid diffusion of AI capabilities coincides with a convergence across firm sizes. In this sense, the post-generative AI period shows not only an increase in the demand for AI skills, but a reshaping of the composition of compensation packages throughout the labor market.

## 5. Discussion

Our analysis provides compelling evidence that non-monetary benefit trends have transformed in the midst of a new era of recruitment of AI talent. While prior research has highlighted the wage premium associated with AI roles (Alekseeva et al., 2021; Bone et al., 2025; Green & Lamby, 2023), our study reveals that firms go beyond monetary offerings with non-monetary benefits — particularly those signaling flexibility, inclusiveness, and support for work-life balance. Importantly, this pattern is not confined to large firms but has become visible across organizations of different sizes alongside the emergence of generative AI and broadened competition for AI talent throughout the labor market.

We find that AI-related job postings are significantly more likely to advertise benefits such as remote work options, an inclusive workplace culture, and parental leave than comparable non-AI vacancies. These benefits — commonly associated with employee well-being and organizational values — have become increasingly prevalent since 2020 and accelerated following the introduction of generative AI in late 2022. In particular, the largest gaps appear for remote work and inclusive workplaces, the least costly entitlements with clear signals, while paid leave and tuition assistance show no AI premium. This may suggest that firms are responding not only to persistent skill shortages but also to evolving worker preferences and changing expectations regarding job quality. These findings are consistent with evidence that workers weigh a broad mix of non-wage conditions when choosing between jobs, including flexibility, autonomy and paid time off (Maestas et al., 2023). It also reflects findings that workers look for employers whose mission and values lend their work meaning (Cassar

& Meier, 2018), and they respond to signals of a firm's social responsibility (Burbano, 2016).

The firm-size results show notable changes in recent years. While S&P 500 firms consistently exhibit the highest demand for AI talent, SMEs show the strongest increase in AI wage premia following the introduction of generative AI. They are also more likely to offer a premium for benefits such as remote work, inclusive workplace, parental leave, and health and well-being initiatives for AI roles. Most of these benefits are categorized as less costly and values-oriented; while parental leave is costly, it can signal values and promote work-life balance and gender equity. At the same time, more costly and traditional benefits — including tuition assistance and paid leave — remain an uncommon premium for AI skills among SMEs. This may reflect the continued importance of resource constraints in shaping compensation strategies, though the data do not prove resource constraints directly, so this is only one interpretation. The stronger AI wage premium among SMEs may also partly reflect composition: an AI role may be more different from the typical SME posting than from the typical large-firm posting. The pattern is nonetheless consistent with SMEs entering competition for AI talent they previously did not seek. The makeup of the SME sample may influence the results; for instance, if they are mostly newer, technology-focused firms, this could partially explain the premium; a more specific decomposition of firm types could further detangle this. Still, regression analyses confirm that these relationships persist after controlling for education, industry, experience, and salary, indicating that they reflect strategies beyond only differences in workforce composition.

Our analysis also extends the literature by showing that the relationship between wages and non-monetary benefits has evolved substantially following the emergence of generative AI. Prior to 2022, complementarity between wages and benefits was largely confined to resource-rich S&P 500 firms, while SMEs more frequently relied on non-monetary benefits as partial substitutes for higher salaries. Following the introduction of generative AI, however, these differences narrow considerably, and complementarity between wages and benefits can be found across firm types. This convergence is consistent with intensified competition for AI-skilled workers for firms across the size distribution, and a change in which firms post AI vacancies. The notable exception is remote work, which increasingly exhibits substitution effects across all firm types, likely reflecting the benefit as a highly coveted one that workers value enough to trade for wages (Mas & Pallais, 2017; Cullen et al., 2025). As complementarity weakened for S&P 500 firms after 2022, other labor-market forces could explain the phenomenon. For instance, factors may have reduced the pressure on these firms to combine pay with benefits, such as layoffs in 2022 and 2023 in the technology sector. Expected productivity gains from AI may also lower firms' expected needs for human capital. Meanwhile, SMEs notably multiplied their search for AI skills in this period and may advertise values-oriented benefits to stand out. Nonetheless, the data only captures advertised benefits, not the supply of talent or motivations behind firms' advertising choices.

Several mechanisms and alternative explanations could explain why certain roles offer more non-monetary benefits. The first is competition for scarce AI talent, which causes firms to advertise more compelling benefits. However, we do not directly observe talent scarcity or employers' motivations, so the analysis cannot establish that competition caused the differences in benefit advertising. Second, the nature of work can influence benefits; for instance, remote work and inclusive workplaces can be more common across data and technology occupations, so this can influence the gap between AI and non-AI roles. Nonetheless, adding occupation controls does not change most estimates, and even among computer and mathematical occupations, AI postings advertise benefits more often than non-AI postings. Third, firms that invest more in digital skills may be productive, have more resources, and employ more people (Jensen & Schneider, 2026). However, SMEs, who may be less productive, offer a premium often at higher levels than S&P 500 firms. There is also a positive, significant difference in the prevalence of all benefits for AI roles compared to non-AI roles both within firms and occupations. This suggests that differences in firms and role types themselves do not fully explain the results, although unobserved vacancy-level differences may remain. Fourth, broader shifts in the U.S. labor market may contribute, such as the normalization of remote work after the pandemic and changes in diversity hiring policy. Though our findings only indicate associations, these alternatives offer interesting avenues for future study to establish causal evidence.

## 6. Contributions and Implications

These findings have several broader implications. First, the study extends findings on AI premia for wages to non-monetary benefits. It shows that the gap in advertised benefits for AI roles varies by firm size and benefit type, and is clearest among SMEs and large firms rather than the S&P 500. It also documents a change after 2022 in how benefits and wages are related within AI postings, with decreased differences across firm sizes. Further, it contributes to literature on recruitment signaling by showing the AI premium is largest for benefits that function most as signals. Benefits that communicate flexibility or organizational values display some of the largest differences between AI and non-AI postings, while widely provided benefits generally distinguish the postings less clearly. Although the analysis does not measure how applicants respond to these signals, it shows that the information employers include in advertisements varies systematically with skill requirements, benefit type, and firm category.

For employers, the results show the overall landscape of compensation, both monetary and non-monetary, for AI roles, alongside increased demand for AI skills across all firm sizes. They highlight the importance of considering advertised benefits as part of the overall compensation package communicated to potential applicants. Meanwhile, they help job seekers evaluate what they are likely to find in recruitment. The study additionally demonstrates which benefits were traditionally substituted for pay and in which firms, and how these trends have changed in recent years. Given the important signaling role of job advertisements (Backhaus, 2004; Rynes et al., 1991), organizations can assess whether they clearly communicate differentiating benefits, particularly those related to flexibility, work-life balance, and organizational culture, considering which benefits are currently used most to distinguish AI vacancies. However, because job postings only reflect what employers offer, not what workers want, these patterns should not be read directly as evidence of worker preferences. In labor markets with search frictions, the combination of wages and amenities that firms offer can differ substantially from how workers actually value those amenities (Hwang et al., 1998).

For policymakers and researchers, the methodology using online job advertisements is a valuable approach to monitor advertised job amenities and how these vary across occupations, skill requirements, and firm categories. The trends in benefit offerings may also have important implications for who companies attract and ultimately hire. Remote work benefits in particular are shown to increase the diversity and experience of applicants, including an increase in women and minorities (Hsu & Tambe, 2025). At the same time, the concentration of high-quality job amenities among AI-intensive occupations may widen inequalities in job quality across the labor market. Further evidence on outcomes across role types such as applications, hires, and retention would be relevant to examine realized impacts on workforce inequalities.

## 7. Limitations

There are several limitations to consider in interpreting our findings. First, the design is largely descriptive, and while we attempt to control for various factors, we cannot fully separate the temporal changes from external forces. The U.S. has seen broader shifts in remote-work and diversity-related hiring practices during this period, which can be seen, for instance, in the steep drop in inclusive workplace benefits listed in 2025. The yearly fixed effects in the regression models control for some of this variability, and the AI advantage in advertised inclusive culture narrows but remains positive through 2025 — and increases for SMEs — indicating that the pattern is not solely an artifact of political changes. Nonetheless, our estimates should be read as associations rather than causal effects. Similarly, many other factors and events occurred pre- and post-2022, and therefore, no causal link can be drawn between the release of ChatGPT in 2022 and the change in benefits.

Second, we observe benefits as advertised in job postings, not as provided; thus, there can be roles where benefits are assumed or listed elsewhere, but not explicitly stated in the job description. This could partially explain why large firms like those in the S&P 500 may not show a substantial premium. They may either expect applicants to assume benefits, offer these across the board for employees, or post them separately on their website. Relatedly, the wage regressions rely on postings with salary information, which may not be representative of all vacancies.

Third, we proxy firm size using posting counts within our sample, which conflates size with hiring intensity and coverage, though our results are largely robust to alternative thresholds. As several of our findings are determined by this firm size assumption, a different classification for firm size could modify the strength of effects for small and large firms, though S&P 500 results would stay consistent. Both SMEs and large firms as categorized here show significant premia for the majority of benefits, while S&P 500 firms are the ones that drive a different pattern.

Fourth, our complementarity and substitution estimates should not be read as direct trade-offs between pay and benefits. Well-paid jobs and benefit-rich jobs may cluster in the same firms because of underlying differences in firm productivity and labor-market frictions (Mas, 2025). Controlling for occupation helps account for variation in job types but does not completely rule it out. A positive interaction between AI wage premia and a benefit may partly reflect which firms offer the benefit, but not intentional decisions to combine them. Future studies could explore firm motivations behind combining, or substituting, salary and job benefits and promoting them in job descriptions to establish an intention to compete for scarce talent. Lastly, we do not directly measure firm resource constraints or actual competition for AI talent. Linking firm-level data, employer surveys, or event-study designs would be valuable directions for future work to help distinguish the above explanations.

## 8. Conclusion

Taken together, these findings show that the emergence of generative AI has been accompanied by both reshaped demand for AI skills and the compensation strategies firms employ to recruit them. Our findings suggest that demand for AI talent has been increasingly widespread across firms of all sizes. Rather than competing exclusively through salaries, organizations differentiate themselves through multidimensional compensation packages whose composition varies with firm size. These non-monetary benefits vary on cost and what they signal to applicants, and can attract different talent depending on how they value such benefits. AI postings are more likely to offer remote work, inclusive workplace culture, and parental leave, a pattern which is most prevalent in SMEs and large firms, and not driven by S&P 500 companies. SMEs have recently seen an important increase in wage premia as well, outpacing larger firms. After generative AI became widely available, the way firms combine wages and benefits became more similar across firm sizes, where previously they followed expected patterns of substitution and complementarity. As AI skills come to be increasingly valued, the new landscape of monetary and non-monetary benefits will be essential for shaping who employers attract and what benefits skilled talent expects.

# Appendix

| | Tuition Assistance | | | | | Paid Leave | | | | |
|---|---|---|---|---|---|---|---|---|---|---|
| | (1) | (2) | (3) | (4) | (5) | (1) | (2) | (3) | (4) | (5) |
| AI Role | −0.033* | −0.035* | −0.106* | −0.153* | −0.107* | −0.197* | −0.135* | −0.126* | −0.134* | −0.044* |
| | (0.010) | (0.010) | (0.015) | (0.015) | (0.015) | (0.007) | (0.007) | (0.010) | (0.010) | (0.010) |
| **Experience (ref. None Listed)** | | | | | | | | | | |
| 0 years | | 0.154* | 0.159* | 0.140* | 0.141* | | 0.063* | −0.055* | −0.061* | −0.053* |
| | | (0.006) | (0.009) | (0.009) | (0.009) | | (0.004) | (0.006) | (0.006) | (0.006) |
| 1-2 years | | 0.074* | 0.005 | 0.008 | −0.001 | | 0.184* | 0.170* | 0.171* | 0.149* |
| | | (0.003) | (0.005) | (0.005) | (0.005) | | (0.002) | (0.003) | (0.003) | (0.003) |
| 3-5 years | | 0.076* | 0.094* | 0.088* | 0.111* | | 0.093* | 0.207* | 0.206* | 0.182* |
| | | (0.003) | (0.005) | (0.006) | (0.006) | | (0.002) | (0.004) | (0.004) | (0.004) |
| 6-10 years | | 0.049* | −0.001 | −0.034* | 0.017 | | −0.104* | 0.053* | 0.047* | 0.046* |
| | | (0.005) | (0.009) | (0.009) | (0.009) | | (0.004) | (0.006) | (0.006) | (0.006) |
| 11-20 years | | 0.048* | −0.168* | −0.273* | −0.205* | | −0.191* | −0.184* | −0.202* | −0.175* |
| | | (0.013) | (0.021) | (0.021) | (0.021) | | (0.009) | (0.014) | (0.014) | (0.014) |
| **Education (ref. No Education Listed)** | | | | | | | | | | |
| High school or GED | | 0.466* | 0.582* | 0.570* | 0.572* | | 0.235* | 0.289* | 0.287* | 0.266* |
| | | (0.003) | (0.005) | (0.005) | (0.005) | | (0.002) | (0.003) | (0.003) | (0.003) |
| Associate degree | | 0.613* | 0.578* | 0.575* | 0.508* | | 0.322* | 0.351* | 0.350* | 0.329* |
| | | (0.005) | (0.008) | (0.008) | (0.008) | | (0.003) | (0.006) | (0.006) | (0.006) |
| Bachelor's degree | | 0.438* | 0.435* | 0.419* | 0.423* | | 0.055* | 0.206* | 0.203* | 0.229* |
| | | (0.003) | (0.006) | (0.006) | (0.006) | | (0.002) | (0.004) | (0.004) | (0.004) |
| Master's degree | | 0.258* | 0.325* | 0.327* | 0.309* | | 0.019* | 0.125* | 0.125* | 0.118* |
| | | (0.008) | (0.013) | (0.013) | (0.013) | | (0.005) | (0.009) | (0.009) | (0.009) |
| Ph.D. or professional degree | | −0.303* | −0.301* | −0.288* | −0.238* | | −0.054* | 0.042* | 0.043* | −0.015 |
| | | (0.015) | (0.024) | (0.024) | (0.025) | | (0.008) | (0.013) | (0.013) | (0.014) |
| log(Salary) | | | 0.157* | 0.127* | 0.017* | | | −0.077* | −0.080* | −0.151* |
| | | | (0.005) | (0.005) | (0.005) | | | (0.003) | (0.003) | (0.003) |
| S&P 500 Firm | | | | 0.599* | 0.598* | | | | 0.101* | 0.107* |
| | | | | (0.006) | (0.006) | | | | (0.005) | (0.005) |
| const | −3.670* | −4.002* | −5.759* | −5.420* | −3.949* | −1.846* | −2.157* | −0.929* | −0.887* | −0.072 |
| | (0.050) | (0.051) | (0.096) | (0.097) | (0.101) | (0.024) | (0.025) | (0.047) | (0.047) | (0.051) |
| Year FE | Yes | Yes | Yes | Yes | Yes | Yes | Yes | Yes | Yes | Yes |
| Industry FE | Yes | Yes | Yes | Yes | Yes | Yes | Yes | Yes | Yes | Yes |
| State FE | No | Yes | Yes | Yes | Yes | No | Yes | Yes | Yes | Yes |
| SOC2 FE | No | No | No | No | Yes | No | No | No | No | Yes |
| Observations | 10.0M | 10.0M | 3.6M | 3.6M | 3.6M | 10.0M | 10.0M | 3.6M | 3.6M | 3.6M |
| Pseudo $R^2$ | .06 | .07 | .08 | .08 | .09 | .06 | .07 | .06 | .06 | .07 |

Note: $^*p < 0.01$. Standard errors in parentheses.

**Table A1: Step-wise logistic regression of benefit provision across roles.** Log-odds coefficients from logistic regressions estimating the probability that a job posting advertises each of the six non-monetary benefits. The unit of observation is a job posting (2018–2025). AI Role is an indicator equal to one if the posting is classified as an AI role.

| | Tuition Assistance | Paid Leave | Health & Wellbeing | Parental Leave | Inclusive Workplace | Remote Work |
|---|---|---|---|---|---|---|
| **Panel A: Pooled 2018–2025** | | | | | | |
| AI Role ($\beta_1$) | 0.274* | 0.277* | 0.269* | 0.265* | 0.258* | 0.292* |
| | (0.002) | (0.003) | (0.002) | (0.002) | (0.002) | (0.002) |
| Benefit ($\beta_2$) | 0.027* | −0.012* | 0.027* | 0.067* | 0.059* | 0.164* |
| | (0.001) | (0.000) | (0.001) | (0.001) | (0.001) | (0.001) |
| AI Role × Benefit ($\beta_3$) | −0.037* | −0.018* | 0.005 | 0.017* | 0.026* | −0.140* |
| | (0.006) | (0.004) | (0.007) | (0.005) | (0.004) | (0.005) |
| AI premium with benefit ($\beta_1+\beta_3$) | 0.238 | 0.258 | 0.274 | 0.282 | 0.285 | 0.152 |
| | (0.005) | (0.003) | (0.006) | (0.005) | (0.004) | (0.004) |
| Observations | 3.5M | 3.5M | 3.5M | 3.5M | 3.5M | 3.5M |
| AI postings | 45K | 45K | 45K | 45K | 45K | 45K |
| $R^2$ | .45 | .45 | .45 | .45 | .45 | .45 |
| **Panel B: Through 2022** | | | | | | |
| AI Role ($\beta_1$) | 0.219* | 0.232* | 0.217* | 0.216* | 0.217* | 0.221* |
| | (0.004) | (0.005) | (0.004) | (0.004) | (0.004) | (0.004) |
| Benefit ($\beta_2$) | 0.026* | −0.014* | 0.012* | 0.044* | 0.037* | 0.208* |
| | (0.001) | (0.001) | (0.001) | (0.002) | (0.001) | (0.001) |
| AI Role × Benefit ($\beta_3$) | −0.023 | −0.047* | −0.005 | 0.004 | −0.012 | −0.093* |
| | (0.014) | (0.008) | (0.014) | (0.016) | (0.010) | (0.010) |
| AI premium with benefit ($\beta_1+\beta_3$) | 0.196 | 0.184 | 0.212 | 0.220 | 0.204 | 0.129 |
| | (0.013) | (0.007) | (0.014) | (0.015) | (0.009) | (0.009) |
| Observations | 1.7M | 1.7M | 1.7M | 1.7M | 1.7M | 1.7M |
| AI postings | 11K | 11K | 11K | 11K | 11K | 11K |
| $R^2$ | .36 | .36 | .36 | .36 | .36 | .37 |
| **Panel C: Post-2022** | | | | | | |
| AI Role ($\beta_1$) | 0.281* | 0.282* | 0.275* | 0.273* | 0.264* | 0.299* |
| | (0.003) | (0.003) | (0.003) | (0.003) | (0.003) | (0.003) |
| Benefit ($\beta_2$) | 0.023* | −0.011* | 0.034* | 0.074* | 0.067* | 0.128* |
| | (0.001) | (0.001) | (0.001) | (0.001) | (0.001) | (0.001) |
| AI Role × Benefit ($\beta_3$) | −0.047* | −0.018* | −0.003 | −0.007 | 0.019* | −0.132* |
| | (0.006) | (0.005) | (0.007) | (0.006) | (0.005) | (0.005) |
| AI premium with benefit ($\beta_1+\beta_3$) | 0.234 | 0.263 | 0.271 | 0.267 | 0.283 | 0.168 |
| | (0.006) | (0.004) | (0.007) | (0.005) | (0.004) | (0.005) |
| Observations | 1.8M | 1.8M | 1.8M | 1.8M | 1.8M | 1.8M |
| AI postings | 34K | 34K | 34K | 34K | 34K | 34K |
| $R^2$ | .46 | .46 | .46 | .46 | .46 | .46 |

Note: OLS of log(salary) on AI role, benefit, and their interaction with education, experience, industry (NAICS3), state, and year fixed effects; HC1 robust standard errors in parentheses. *$p < 0.01$.

**Table A2:** Wage–benefit interactions for AI roles. OLS estimates of equation (2) by benefit, for the pooled sample (Panel A) and before/after 2022 (Panels B–C). HC1 robust standard errors in parentheses; *$p < 0.01$.

| Variable | Percentage |
|---|---|
| AI Role (133,369 postings) | 1.3% |
| **Benefits (% offering each benefit)** | |
| Paid Leave | 28.4% |
| Inclusive Workplace | 10.6% |
| Tuition Assistance | 9.2% |
| Remote Work | 6.6% |
| Health and Wellbeing | 5.1% |
| Parental Leave | 4.7% |
| **Experience Requirements** | |
| None Listed | 54.6% |
| 1-2 years | 20.4% |
| 3-5 years | 15.4% |
| 6-10 years | 5.4% |
| 0 years | 3.4% |
| 11-20 years | .8% |
| **Education Requirements** | |
| No Education Listed | 46.4% |
| High school or GED | 24.8% |
| Bachelor's degree | 20.7% |
| Associate degree | 5.0% |
| Master's degree | 2.1% |
| Ph.D. or professional degree | .9% |

*Note:* Sample includes 9,999,673 job postings from 2018-2025.

**Table A3. Summary of the variables in the main regression model.**

| Benefit | Precision | Recall | F1 |
|---|---|---|---|
| Parental Leave | 0.938 | 0.962 | 0.950 |
| Paid Leave | 0.930 | 0.962 | 0.946 |
| Tuition Assistance | 0.893 | 0.981 | 0.935 |

**Table A4. Agreement between keyword-based benefit labels and Lightcast-labeled benefit fields.** Precision, recall, and F1 scores for keyword-detected benefit labels, treating Lightcast labels as the reference. Based on 9,999,673 U.S. job postings, 2018–2025.

## *Figures*

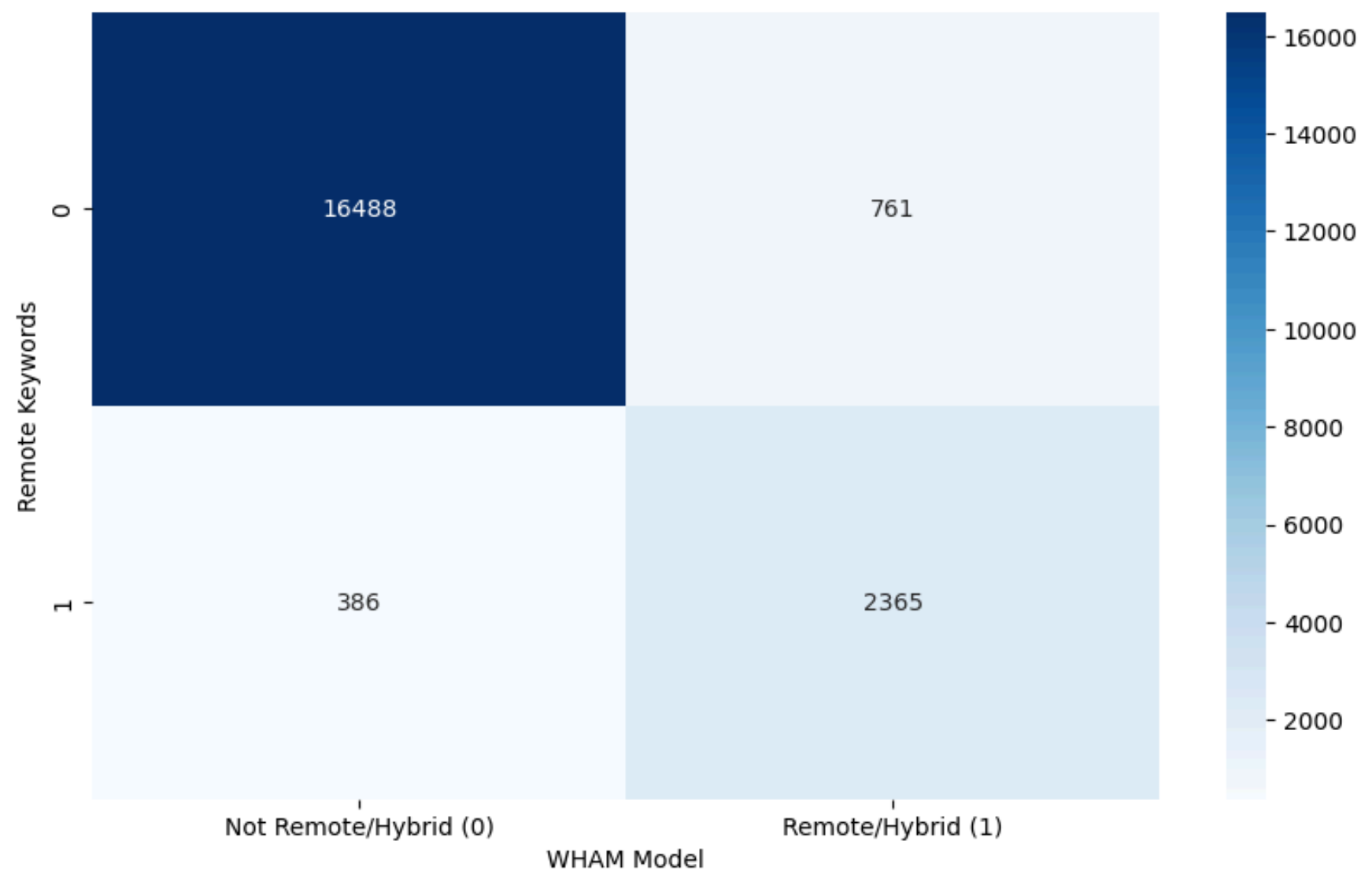

**Figure A1. Validation of remote work classification via cross-comparison with the WHAM model.** This figure presents a cross-validation matrix comparing our keyword-based remote work classification with the WHAM benchmark model (Hansen et al., 2023). The comparison reveals high classification accuracy: 98% of non-remote roles and 76% remote roles are correctly identified using our approach. Overall, our method achieves an accuracy of 94% and an F1 score of 0.805, validating its robustness beyond hand-coded checks.

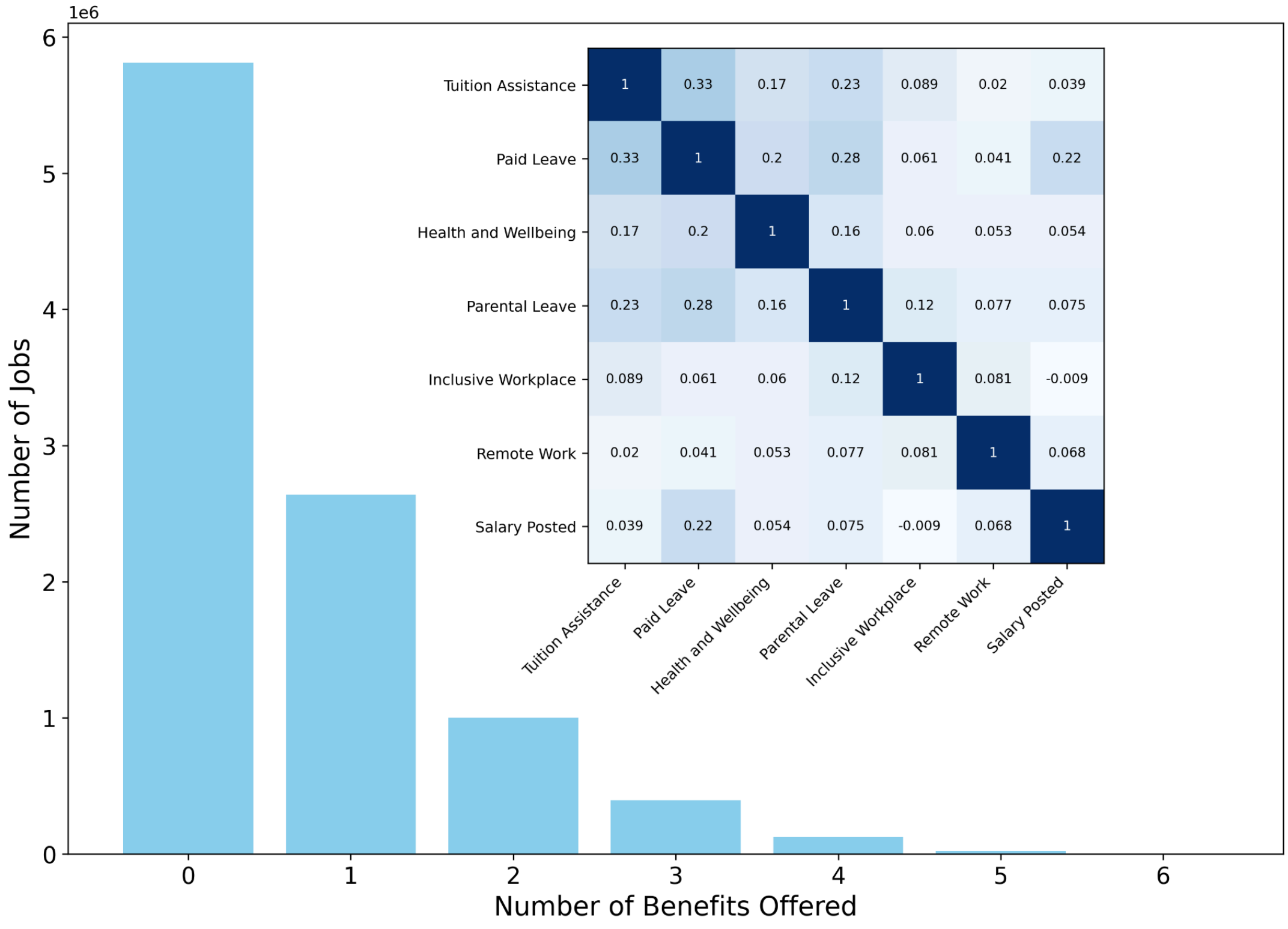

**Figure A2. Distribution of benefit bundling across all job postings.** The figure shows that most job postings offer no non-monetary benefits, and among those that do, the vast majority offer only one. Roles offering two or more benefits are substantially rarer. The inset highlights that the most frequent pairings are paid leave with tuition assistance, and paid leave with parental leave, indicating common co-occurrence patterns.

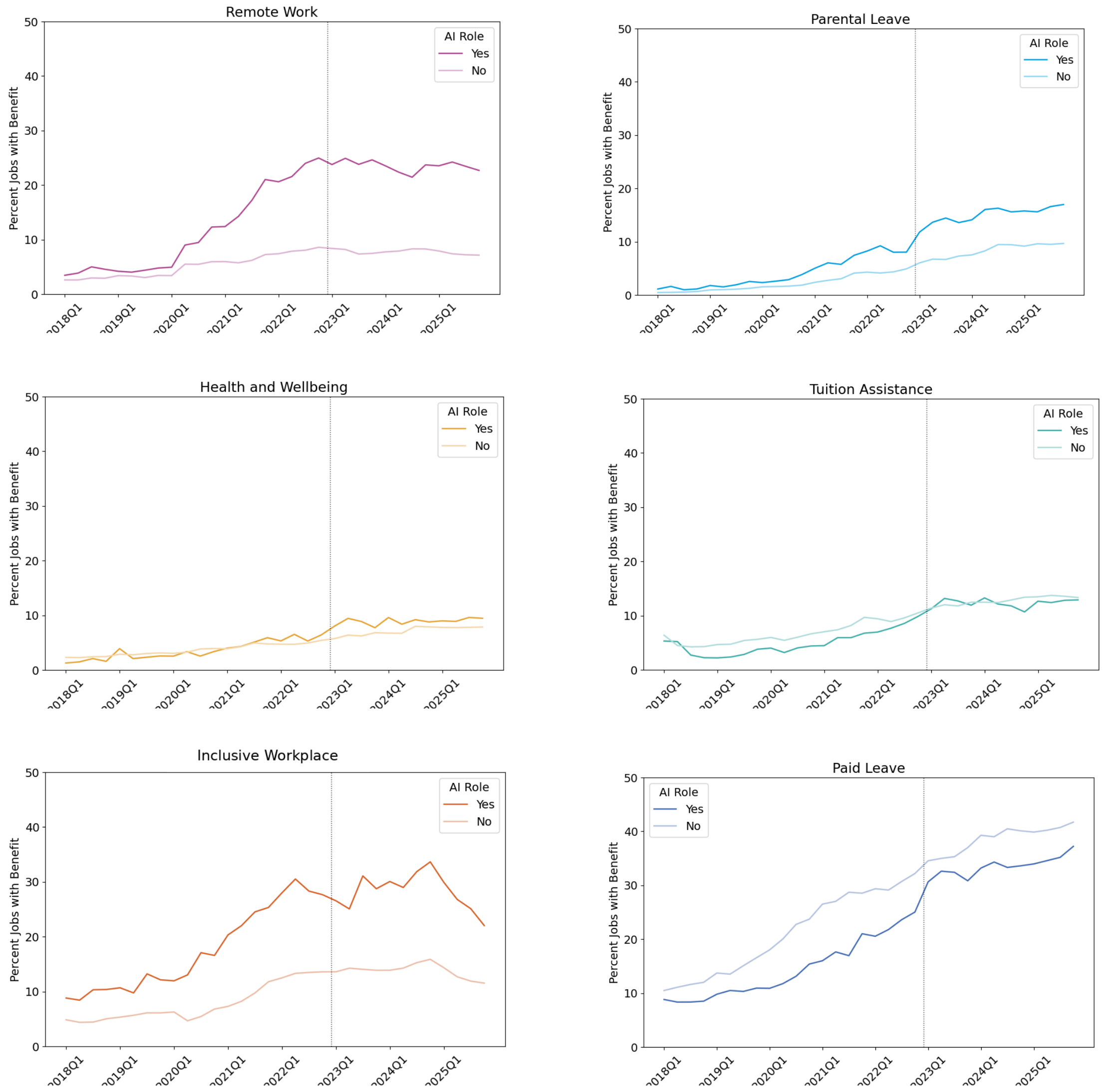

**Figure A3. Comparison of benefit prevalence between AI and non-AI roles.** This figure tracks the relative share of AI and non-AI roles offering specific benefits. A clear divergence emerges for inclusive culture, remote work, and parental leave, with AI roles increasingly offering these perks more frequently. Other benefits move in parallel across role types, though AI roles appear to be catching up with non-AI roles in the provision of paid leave.

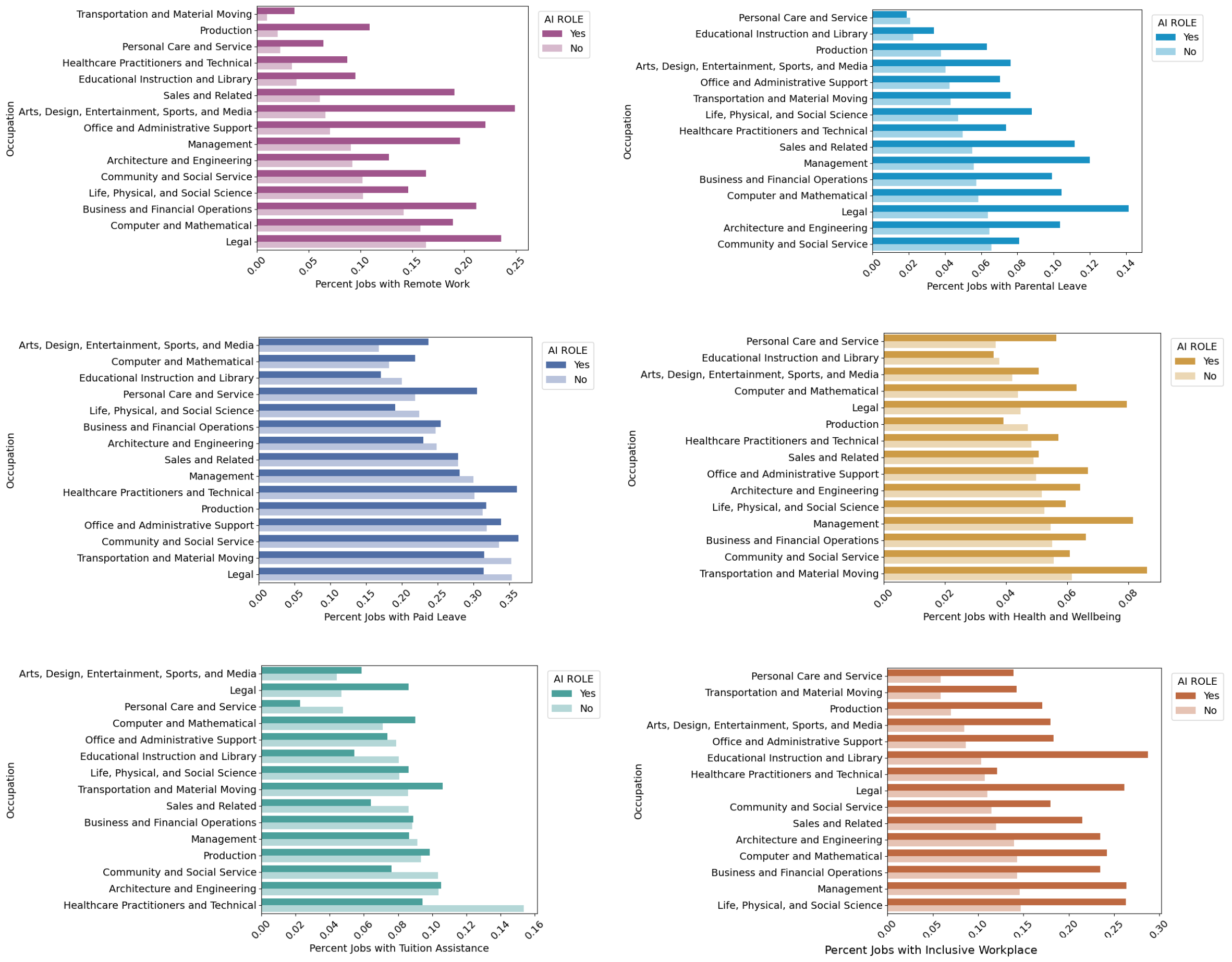

**Figure A4. Benefit prevalence across occupational groups.** This figure explores how non-monetary benefits are distributed across occupations. While some sectoral differences emerge—such as remote work being particularly common in Arts, Design, and Sports and much rarer in physically intensive roles—the overall pattern suggests that the observed benefit gaps between AI and non-AI roles are not driven by a single occupation, but are broadly consistent across the labor market.

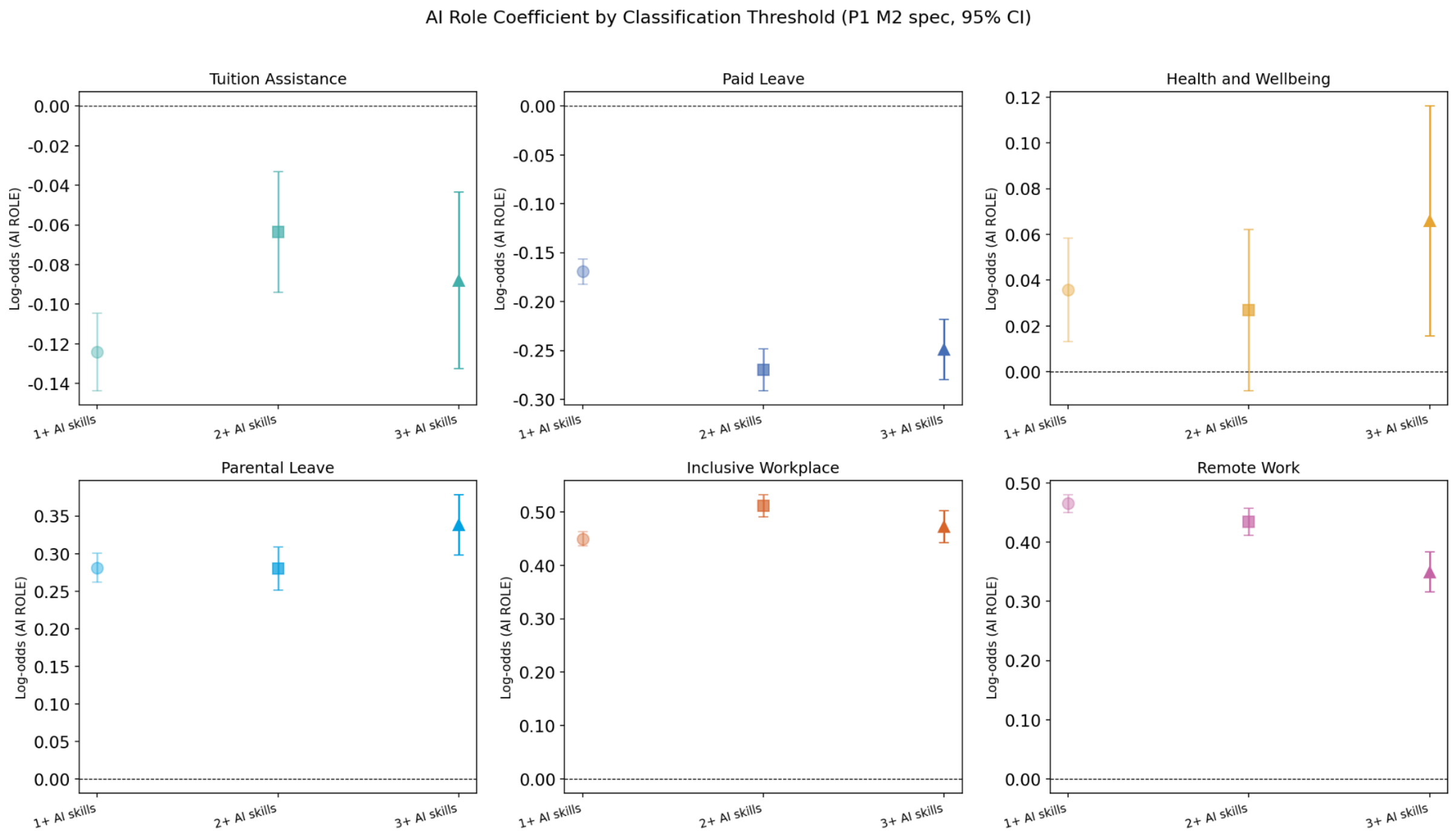

**Figure A5. Robustness check of AI role classification for varying AI skill counts.** We ran the first regression using different thresholds for classifying an AI role, requiring one, two, or three AI skills for a role to be classified as an AI role. Results show that the different thresholds produce similar coefficients, though with increasing error margins due to the increasing sparsity of AI roles with more skills required.

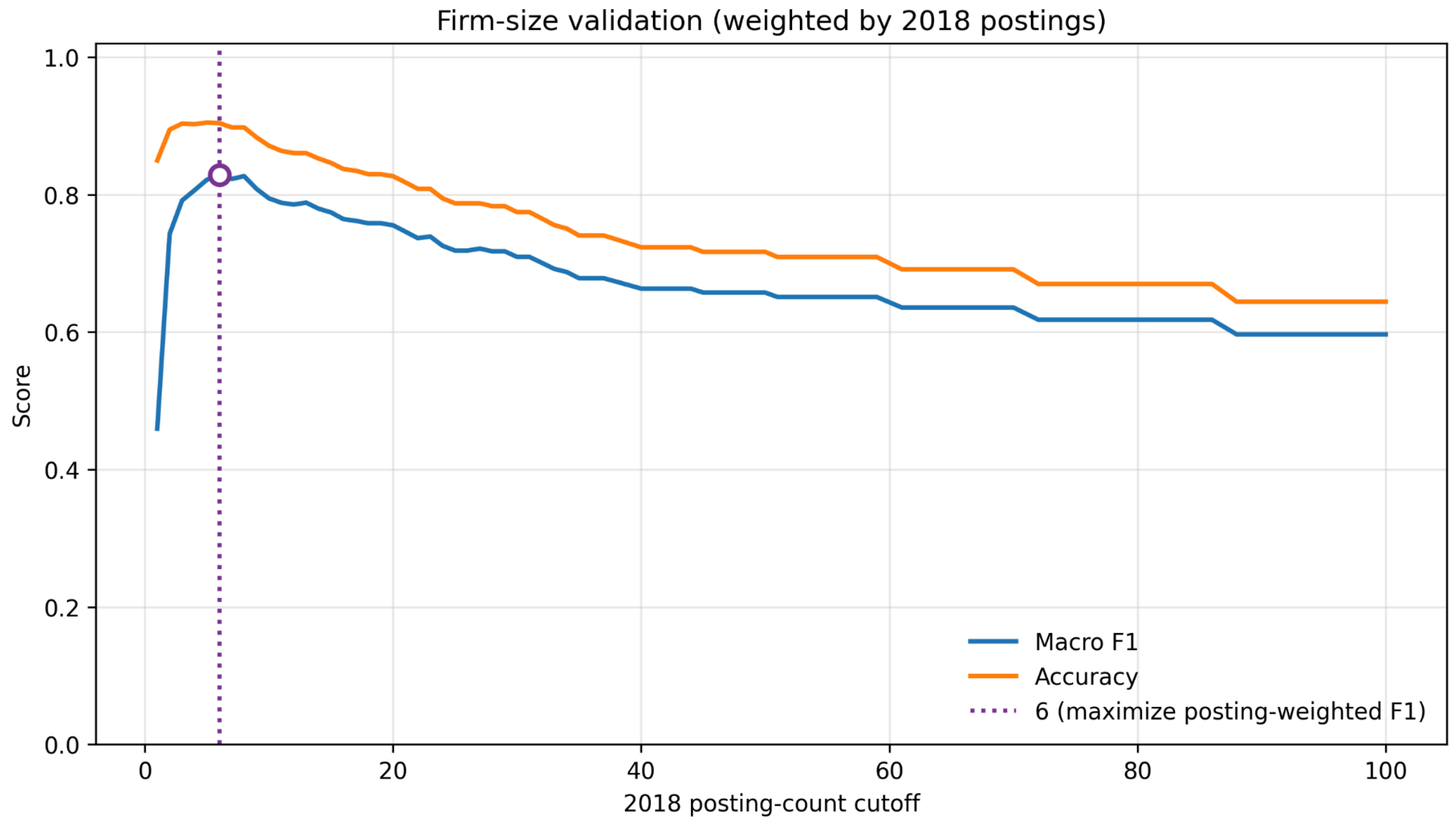

**Figure A6. Firm size threshold performance metrics.** This plot shows the performance metrics of using different post-count cutoff values to divide SMEs and Large employers.

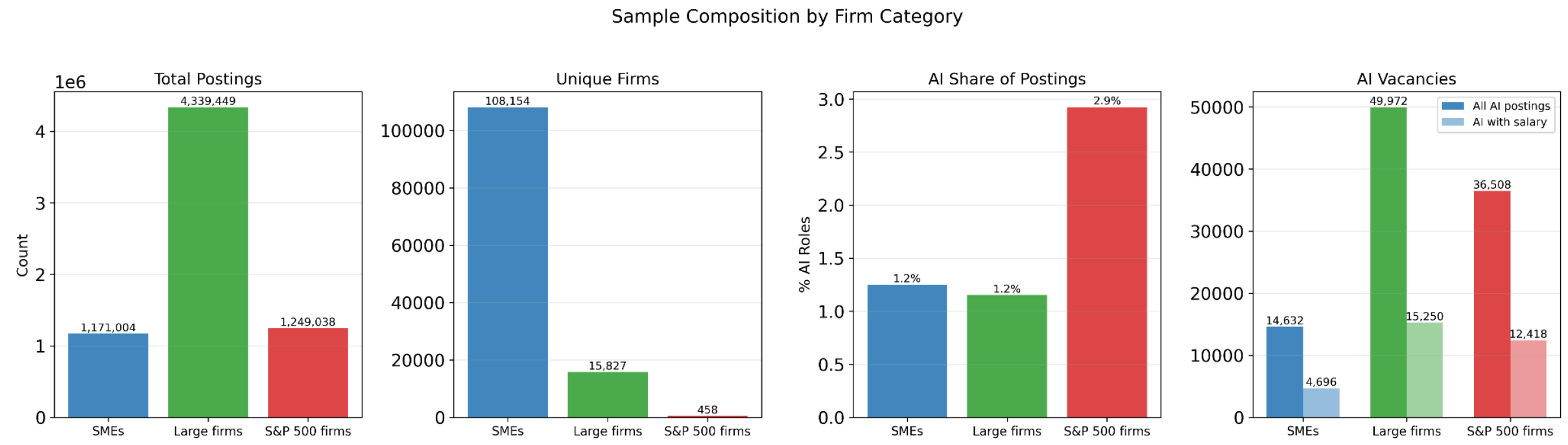

**Figure A7.** Summary of AI role prevalence across firm sizes.

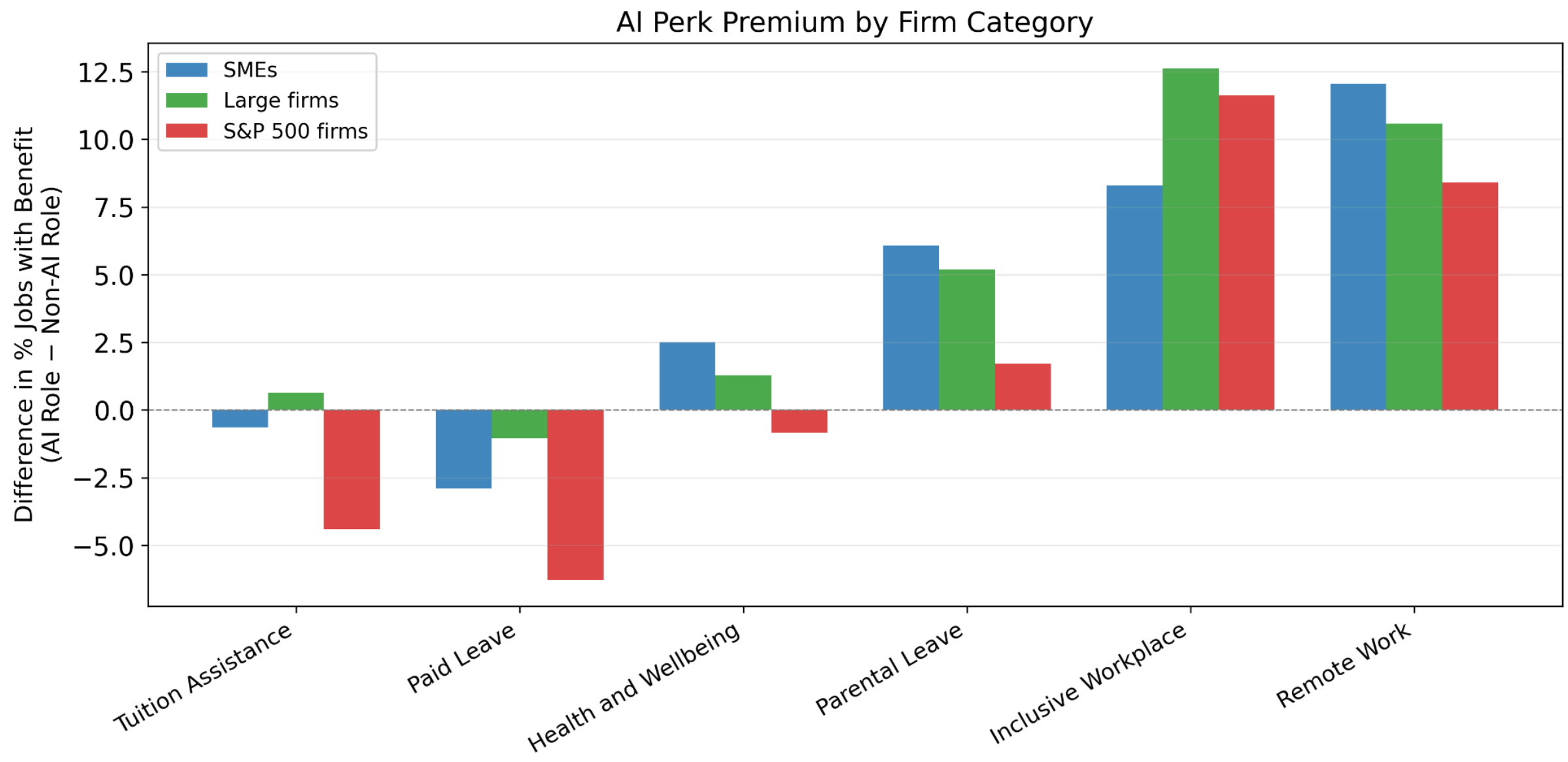

**Figure A8.** Summary of perk prevalence for AI roles and non-AI roles across firm sizes

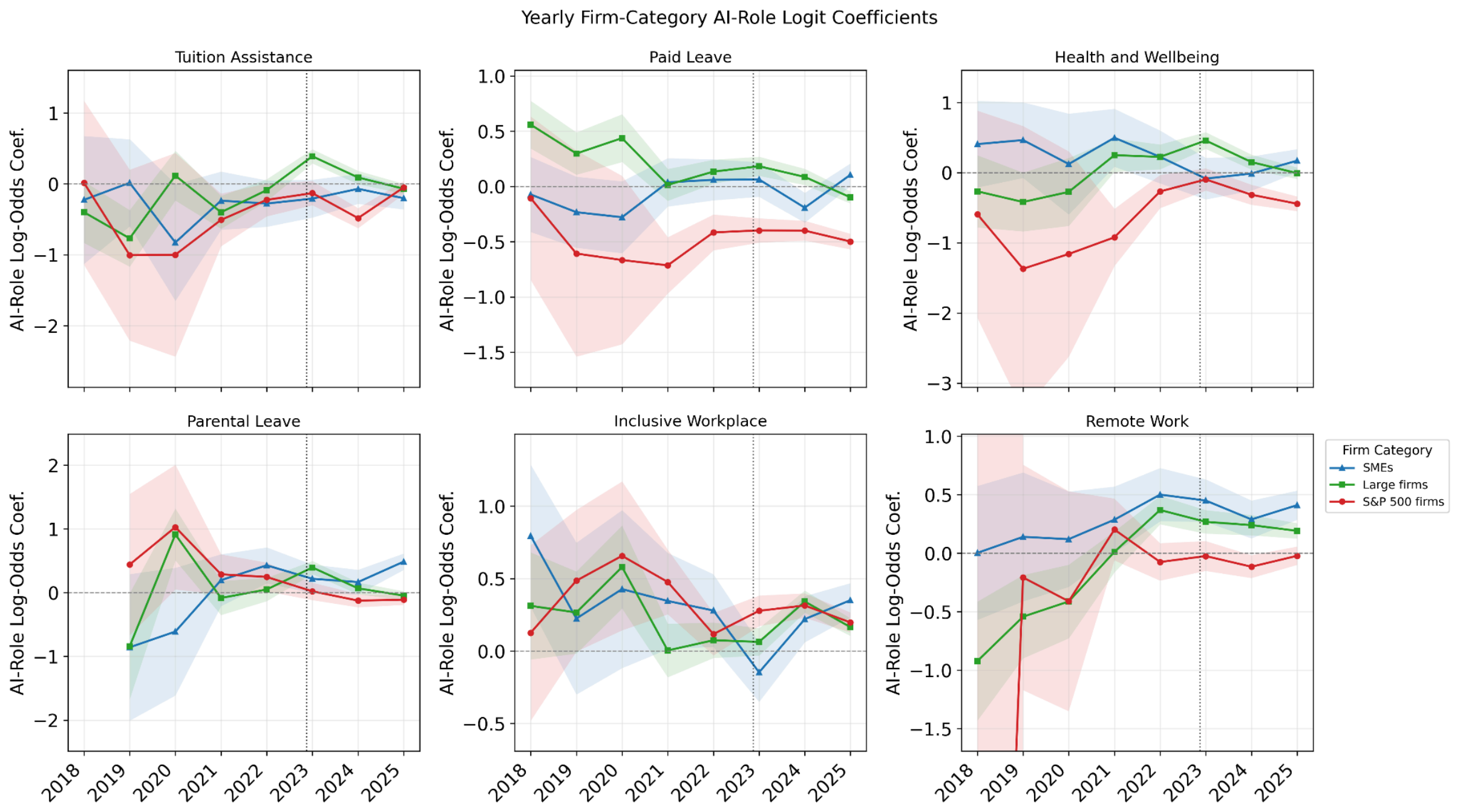

**Figure A9.** Yearly AI role Log-Odds Coefficient for Regression 1.

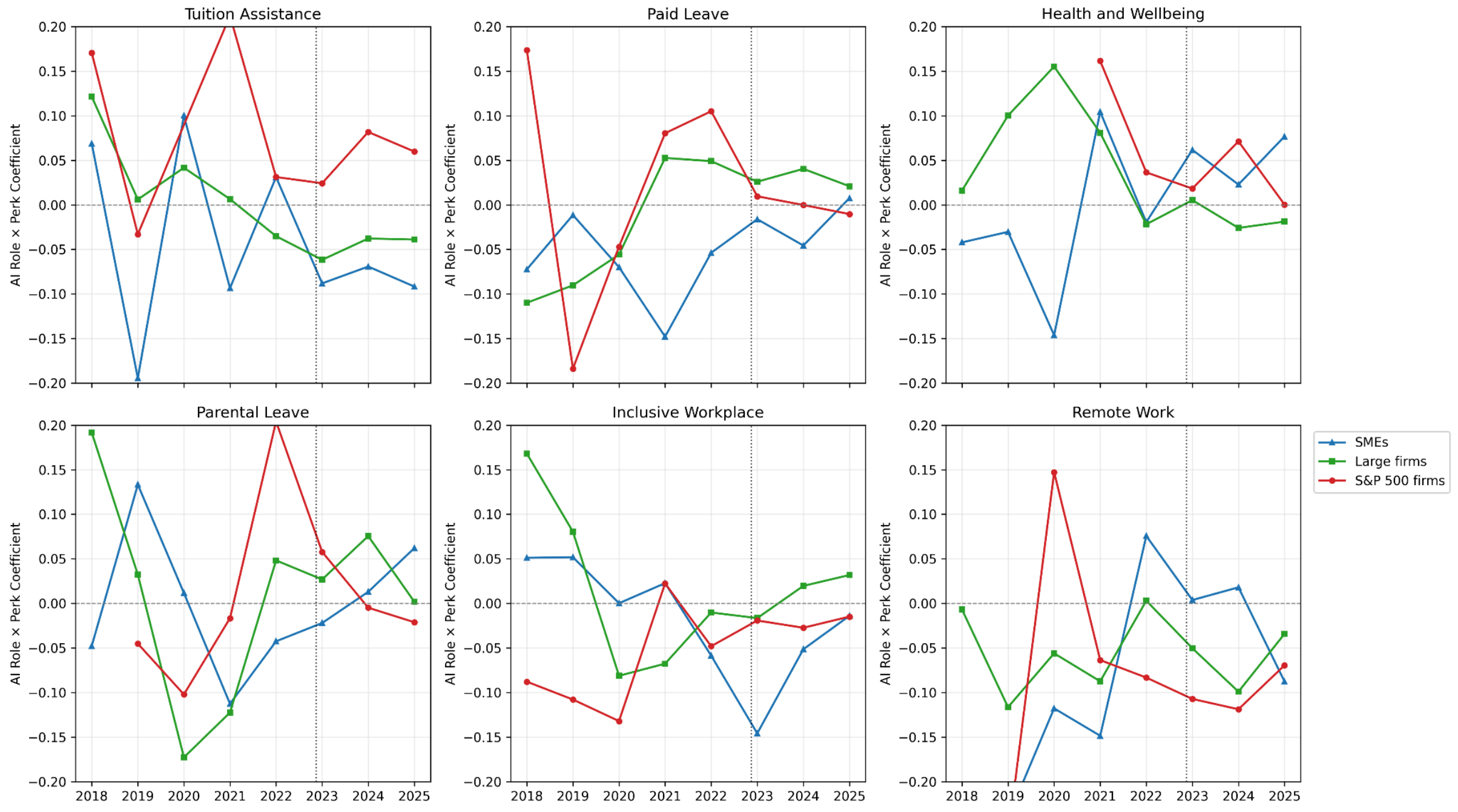

**Figure A10.** Yearly AI Role x Perk Interaction Coefficient for OLS Regression on Wages.

## Differences in benefit prevalence and salaries for AI and non-AI jobs within occupations and firms

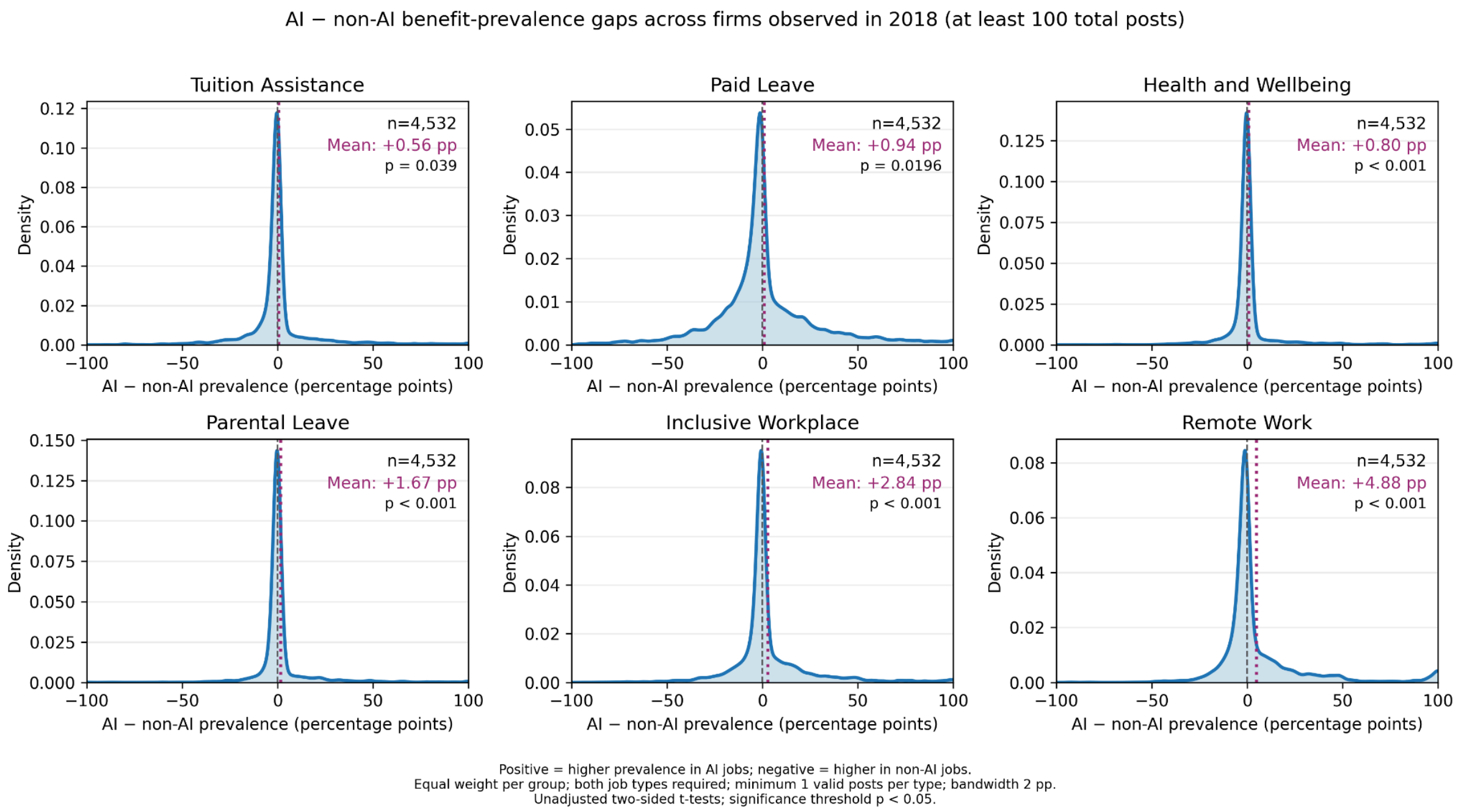

**Figure A11.** Plot showing the difference in benefit prevalence between AI and non-AI jobs within firms. This shows that firms' AI jobs tend to have more benefits when not controlling for any other factors.

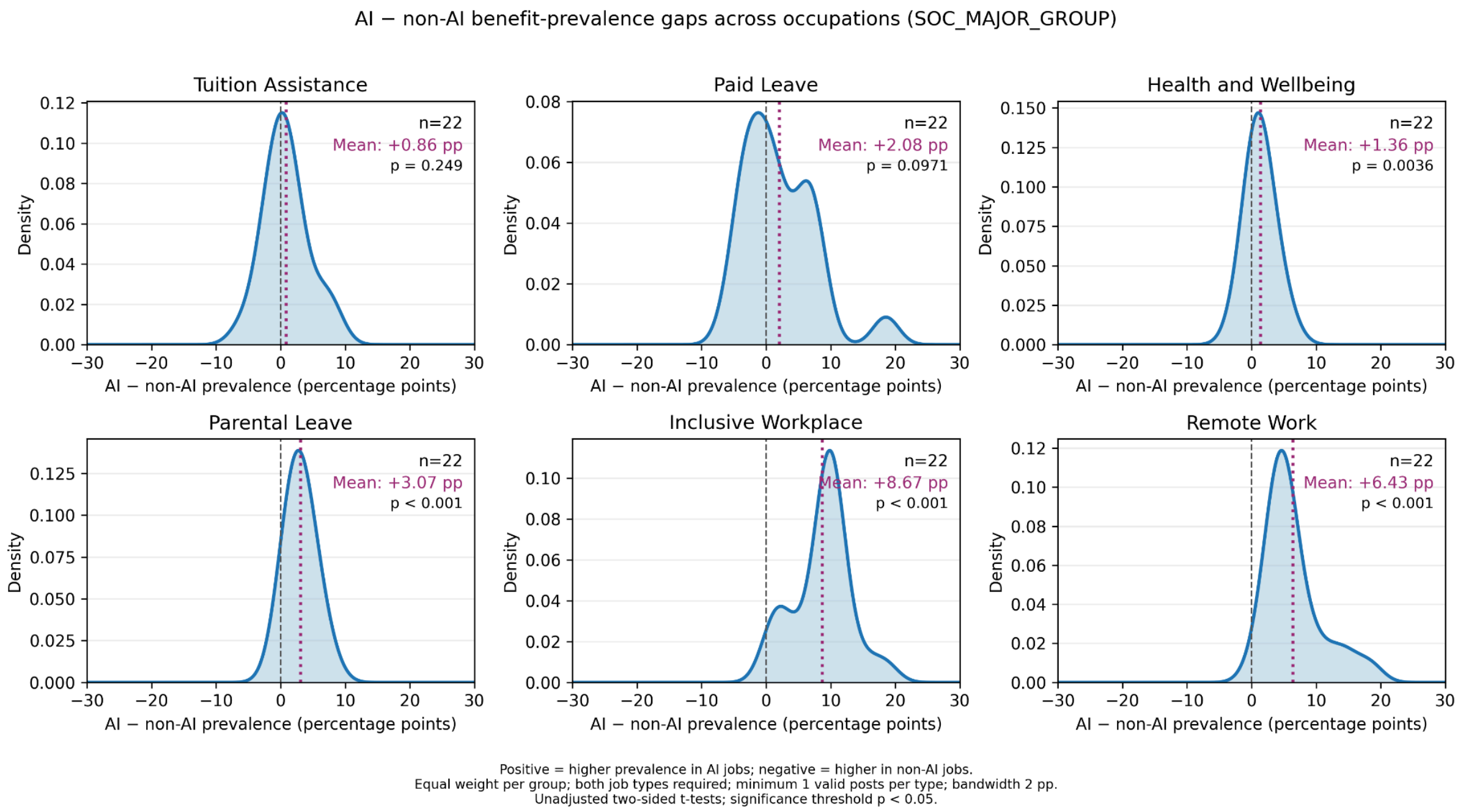

**Figure A12.** Plot showing the difference in benefit prevalence between AI and non-AI jobs within occupations. This shows that occupations' AI jobs tend to have more benefits when not controlling for any other factors.

The above plots indicate that, descriptively, AI jobs tend to be associated with more benefits within firms and within occupation groups. Some extreme but unsurprising examples of this include Walmart which offers remote work in 26% of its AI jobs but only 1% of its non-AI jobs. On the flipside, this also reveals some instances where, presumably the benefits are implicit in the job posting rather than explicitly listed. For instance, only 3% of AI jobs in Amazon include paid leave whereas 47% of their non-AI jobs explicitly mention it. When it comes to occupations, we can see that within some remote-friendly occupations like computer and mathematical workers, the remote work gap is relatively small but persists with 19% of AI jobs offering it while only 16% of non-AI jobs offer it.

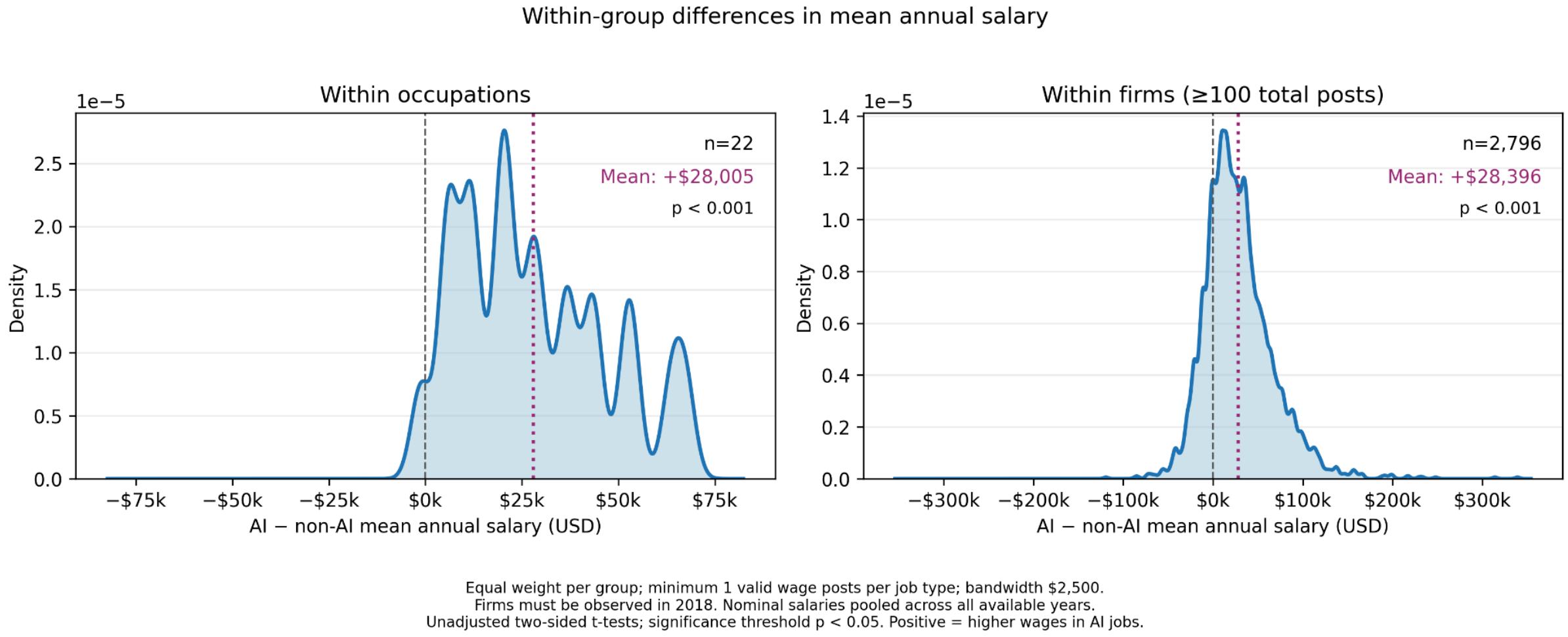

**Figure A13.** Plot showing the difference in annual salaries between AI and non-AI jobs within occupations. This shows that occupations' AI jobs tend to be associated with higher pay within occupations and firms when not controlling for any other factors.

## Keyword inclusion lists

Tuition Assistance: ["education assistance", "tuition reimbursement", "tuition assistance", "education reimbursement"]

Paid Leave: ["paid time off", "PTO", "extra vacation", "flexible vacation", "mental health day", "generous time-off", "generous time off", "paid vacation", "paid holidays", "holiday pay", "vacation days", "company holiday", "sick leave", "vacation time", "paid days off", "paid flexible holidays"]

Health & well-being: ["wellness stipend", "health benefits", "mental health support", "gym membership", "wellness program", "well-being stipend", "wellness programs", "mental health benefits", "employee well-being", "health care benefits"]

Parental Leave: ["parental leave", "family support", "flexible maternity leave", "paternity leave", "maternity leave", "paid family leave", "paid caregiver/parental", "paid parental", "paid new parent leave", "paid bonding leave", "parental bonding leave"]

Inclusive Culture: ["diversity and inclusion", "team culture", "creative freedom", "values-driven", "inclusive culture", "diverse team", "inclusive environment", "value diversity", "diversity, equity", "culture of diversity", "diversity, inclusion", "inclusion and diversity", "commitment to diversity", "diversity, inclusion", "diversity, equity and inclusion", "diversity is respected", "inclusive diversity", "workforce diversity", "embracing diversity", "value diversity", "values diversity", "committed to diversity", "equity and diversity", "promoting diversity", "celebrate diversity", "encourage diversity", "support diversity", "diversity in the workplace", "equity, inclusion"]

Remote Work: ["fully remote", "100% remote", "work from home", "remote role", "work remotely", "remote eligible", "open to remote", "remote work environment", "remote work policy", "remote and onsite", "virtual role", "remote first", "home office", "telecommute", "distributed team", "remote-first", "remote-friendly", "remote options", "virtual work", "work from home", "remote position", "wfh", "Remote - us", "telework", "home office", "(remote)", "remote work flexibility", "remote work: hybrid", "remote: yes", "location: remote", "remote usa", "remote flexibility", "(remote", "- remote", "remote-flexibility", "remote,", "\nremote\n", "\n remote \n", "remotely", "us-remote", "remote work eligible", "remote - united states", "remote - work at home", "can be remote", "remote within the us", "remote in north america", "remote available" ", remote", "remote full-time", "us remote", "#li-remote", "open to remote", "hybrid work", "split time between office and remote", "office and remote options", "partially remote", "hybrid position", "3 days remote", "hybrid workplace", "some remote days", "in-office and remote", "hybrid onsite", "remote or in-office", "work-from-home days", "mix of working in the office and from home", "#li-hybrid", "hybrid remote"]

**Keywords exclusion lists**

Tuition: ["youth leadership development", "provide mentorship", "professional development expertise","planning professional development", "forecasting growth opportunities", "teen leadership development", "develop training programs", "professional development experience", "training programs as required", "providing mentorship", "providing and encouraging mentorship", "implementing training programs", "provide leadership and mentorship", "provide mentorship", "identify growth opportunities", "manage the Symbotic Tuition Reimbursement"]

Health & Well-being: ["health benefits companies", "wellness program coordinator", "familiarity with mental health support", "health benefits administration"]

Parental Leave: ["accredited childcare program", "maui family support services", "teaching assistant - childcare", "experience in childcare", "childcare state licensing", "experience with children"]

Inclusive workplace: ["maintaining a collaborative environment"]

Remote Work: ["not considering remote", "in-office only", "remote monitoring", "remote sensing", "remote access systems", "remote control", "remote diagnostics", "remote delivery", "#li-onsite", "onsite job", "work from home not available", "telework:no", "remote: no", "100% on-site", "work at home option: No", "remotely: no", "remote: n", "remotely: n", "telework: no", "remote: * no", "remotely: * no", "remotely piloted", "data remotely", "remote type on-site", "interviewed remotely", "work remotely: * no", "remote desktop", "must be able to work on-site", "not applicable for 100% remote", "remote testing", "remotely upgrading", "remote usability", "remote site", "remote machine", "remote iot", "no remote", "work remotely no", "remotely:no", "remotely? n", "remotely no", "remote areas", "remote access", "remote position? no", "remotely tucked away", "supporting remote", "remotely sensed"]

## Firm-size Validation Prompt

```
"You classify employers by their true real-world organization size.\n\n"
  "Use the employer name, industry context, and general knowledge of "
  "the employer's real-world scale.\n\n"
  "Return only strict JSON with keys:\n"
  '- label: one of "sme", "large", or "unknown"\n'
  "- confidence: number from 0 to 1\n\n"
  "Definitions:\n"
  'Use "large" for employers that are likely large enterprises, including '
```

```
"national or multinational corporations, public companies, subsidiaries "
"or brands of large parent companies, large chains, large hospital or "
"university systems, government entities, national nonprofits, major "
"staffing firms, or professional-services firms operating across many "
"locations.\n\n"
'Use "sme" for employers that are likely small or medium-sized independent '
"organizations, including local or regional businesses, independent "
"clinics, restaurants, shops, agencies, contractors, schools, nonprofits, "
"or niche private firms without evidence of national scale.\n\n"
'Use "unknown" only when the employer name is missing, generic, ambiguous, '
"confidential, a job-board/staffing placeholder, or otherwise impossible "
"to classify from the name and industry context.\n\n"
'Do not label a firm "unknown" merely because it is unfamiliar. If the '
"employer appears to be an independent local, regional, or niche "
"organization and there is no evidence of national or multinational "
'scale, classify it as "sme".'
```